\documentclass[12pt]{elsarticle}
\pdfoutput=1
\usepackage{xcolor,graphicx}
\usepackage{textcomp}
\usepackage{amsmath}
\usepackage{amssymb}
\usepackage{verbatim}
\usepackage[colorlinks = true,
            linkcolor = cyan,
            urlcolor  = blue,
            citecolor = red,
            anchorcolor = blue]{hyperref}

\makeatletter
\DeclareRobustCommand{\text}{%
  \ifmmode\expandafter\text@\else\expandafter\mbox\fi}
\let\nfss@text\text
\def\text@#1{{\mathchoice
  {\textdef@\displaystyle\f@size{#1}}%
  {\textdef@\textstyle\f@size{#1}}%
  {\textdef@\textstyle\sf@size{#1}}%
  {\textdef@\textstyle \ssf@size{#1}}%
  \check@mathfonts
  }%
}
\def\textdef@#1#2#3{\hbox{{%
                    \everymath{#1}%
                    \let\f@size#2\selectfont
                    #3}}}
\makeatother

\title{A Prototype Neutron Veto for Dark Matter Detectors}
\author{S. Westerdale, E. Shields, and F. Calaprice}
\address{Department of Physics, Jadwin Hall, Princeton University. Princeton NJ 08544}
\date{2015}
\begin{document}
\begin{abstract}
Neutrons are a particularly dangerous background for direct WIMP dark matter searches; their nuclear recoils with the target nucleus are often indistinguishable from nuclear recoils produced by WIMP-nuclear collisions.  In this study, we explore the concept of a liquid scintillator neutron veto detector that would allow direct dark matter detectors to potentially reject neutrons with greater than 99\% efficiency.  Here we outline the construction and testing of a small prototype detector and the potential implications of this technology for future dark matter detectors.
\end{abstract}
\maketitle

\section{Introduction}

The search for dark matter in the form of Weakly Interacting Massive Particles (WIMPS) by direct detection consists of searching for low energy nuclear recoils that are produced by WIMP-nuclear collisions.   The energy of the nuclear recoil is expected to be low, of the order of 100 keV, and the event rates of interest for future experiments could be $\approx$ 1 count/ton/yr, or lower.  Achieving this sensitivity in future WIMP searches will require detectors of large mass and extremely low backgrounds.

Even with careful selection of low background materials for the internal parts of the detector and careful shielding against external backgrounds, natural sources of radioactivity pose a serious challenge.    Neutrons are a particularly dangerous background; a neutron that scatters off a target nucleus in the detector can produce a signal that is identical to that of a WIMP collision.  
Rejecting neutron backgrounds by their multi-hit signature in the active volume or by using self-shielding can be quite effective. However, both of these techniques require very large masses in order to be effective, and using them may greatly reduce the detector's fiducial mass and exposure.

Neutrons can produce a signal in a dark matter detector primarily from two different types of sources: internal, from radioactive contamination of the detector materials, and external, either produced from cosmogenic muons or from the surrounding environment. No amount of passive shielding can guard against neutrons from the former source, since they are produced next to the dark matter detector, typically from either the fission of heavy nuclei like uranium and thorium, commonly found in steel samples and PMTs, or from the $(\alpha,n)$ reaction when $\alpha$-emitters interact with light nuclei. These neutrons will be present regardless of the amount of shielding used, but they can be reduced, though not completely eliminated, through careful material selection. However, a neutron veto would be able to detect these neutrons when they leave the detector and efficiently eliminate these backgrounds. 

External neutrons, on the other hand, can be reduced by the addition of passive shielding. This is especially so for neutrons from the surrounding environment. However, cosmogenic neutrons can be produced at much higher energies of hundreds of MeV when a muon passes by the detector. Muons can produce neutrons either through various processes including spallation on heavy nuclei. These high energy neutrons can travel extremely long distances, and will often have a mean free path of a few meters, making them impractical to block with a passive shield.  However, an active veto system can effectively eliminate these backgrounds by detecting the neutrons when they scatter in the veto, before or after the neutron scatters in the dark matter detector. Additionally, a water Cherenkov detector can be used to detect passing muons that might produce such a neutron. FLUKA simulations of these scenarios performed by~\cite{Empl:2014ih} have shown that these neutrons can be significantly reduced by the use of an external active veto system.

This report describes studies made with a prototype of such a neutron veto detector based on a boron-loaded liquid scintillator. The detector design that we explore is a vessel with 4$\pi$-coverage of a dark matter detector.  The liquid scintillator, as studied here, is an organic solvent with a wavelength shifter.  When a scintillation event occurs in the liquid scintillator, light is produced, which is then shifted to a visible wavelength and collected in photomultiplier tubes (PMTs) surrounding the walls of the detector.  This event can act as a veto for the coincidental presence of a neutron in the dark matter detector.  An earlier Monte Carlo study showed that the efficiency of such a detector to reject neutron backgrounds could be greater than 99\% ~\cite{Wri2010}.  We report here on studies of practical details that affect the overall performance of the detector, particularly the choice of scintillator and reflector for achieving high light yield.  We motivate this study as part of the development of a neutron veto for the DarkSide-50 WIMP detector, an $^{39}$Ar-depleted liquid argon dual-phase Time Projection Chamber (TPC) experiment.

The efficiency for neutron detection in this type of veto detector depends on two factors.  It depends first on the neutron capture cross section of the scintillator.  Certain elements have high cross sections, such as lithium, gadolinium, and boron~\cite{Knoll}.  The scintillator can be loaded with one of these elements in order to increase the probability that a neutron will cause a visible scintillation event after entering the veto. s

Boron can be added to a scintillator like pseudocumene (PC) via trimethyl borate (TMB).  The boron-loaded scintillator captures thermal neutrons with a very high cross section (3838 barns) through one of two channels ~\cite{Wri2010}:
\begin{gather}
 \begin{aligned}
  ^{10}\text{B}+n &\rightarrow\ ^7\text{Li}^* + \alpha \text{ (1471 keV)} &&\text{ (93.7\%)}\\ 
                  & \hspace{23pt} ^7\text{Li}^* \rightarrow ^7\text{Li (839 keV)} + \gamma \text{ (478 keV)} \\
                  &\rightarrow\ ^7\text{Li (1015 keV)} + \alpha \text{ (1775 keV)} &&\text{ (6.4\%)}
 \label{reaction}
 \end{aligned}
\end{gather}

The first channel produces an excited lithium nucleus, which de-excites and produces a $\gamma$ ray that is easily detected.  The rarer second channel, however, gives a smaller signal, since nuclear recoils in the scintillator are quenched to the level of 30--60 keVee or lower (as measured by this report and others~\cite{Wri2010,ScintillatorS1,ScintillatorS2,ScintillatorT}), and are therefore harder to see than the electron recoils produced by $\gamma$ recoils, which are quenched very little.  However, due to the extremely short mean free path of nuclei in the scintillator, the $\alpha$ and $^7$Li nucleus from these two interactions will almost never escape the neutron detector, while the $\gamma$ might. This means that a veto with a threshold low enough to reliably see this $\alpha$ can detect neutrons capturing on the boron nucleus with a very high efficiency.

This second reaction, with its high quenched energy deposits, calls attention to the second factor that affects the veto efficiency.  It is important that the light collection in the veto be as high as possible, in order to detect the light from these low-energy reaction products. The scintillator itself should have a high scintillation yield, to produce the maximum number of photons to start with.  The attenuation length of the scintillator should also be larger than the size of the veto, so that with a good reflecting surfaces, the light can travel far enough to reach the photodetectors.

Complete PMT coverage would not be cost-effective for a large veto, so the walls of the veto are lined with a highly reflective material. This will reduce the probability of a photon being absorbed before reaching a PMT.  Such a reflector must be compatible with and chemically resistant to the scintillator.  It also must maintain its high reflectance when immersed in the liquid and have a low radioactivity.

The PMTs should have a high quantum efficiency, which should be maximized in the same wavelength region as the reflector and the scintillator emission spectra.  

Assuming the veto can efficiently detect neutrons, the biggest source of inefficiency will come from neutrons that capture in the materials of the dark matter detector and do not make it into the veto. Two elements commonly found in dark matter detectors that may capture neutrons, albeit with a relatively low capture cross sections compared to $^{10}$B, include $^{19}$F, which is frequently found in reflectors like Teflon, and $^{56}$Fe, which is found in steel. These two isotopes may capture and produce a 6.6 or 7.6 MeV $\gamma$, respectively~\cite{endf}. With an acquisition window extending $\sim$60 $\mu$s after the neutron was seen in the dark matter detector, some fraction of these neutrons can be recovered by detecting the $\gamma$ produced in the $^{19}$F or $^{56}$Fe capture reaction.

We have built a prototype neutron veto, developing the specific technologies required for a large-scale veto for a dark matter detector.  This technology was first applied to the DarkSide-50 liquid scintillator veto. We discuss the development of the scintillator cocktail, the choice of the reflector, and the light yield of our prototype.  
\section{Scintillator}
The scintillator mixtures we consider consist of three to four distinct parts: a scintillating solvent (PC), a secondary solvent with a high neutron cross section (TMB), a primary wavelength shifter (2,5-diphenyloxazole, or PPO), and in some cases a secondary wavelength shifter. When we used a secondary wavelength shifter, we used 1,4-bis(5-phenyloxazol-2-yl)benzene (POPOP) or 1,4-bis(2-methylstyryl)benzene (bis-MSB).

One important consideration when acquiring the solvents used in the veto is the contamination of $^{14}$C in the solvent. Organic compounds are typically created using a methanol base, which may come from modern biogenic material, typically derived from plant matter. Organic compounds made from plant-based methanol tend to have approximately modern atmospheric levels of $^{14}$C, which $\beta^-$ decays with an endpoint of 156 keV. Measurements of modern atmospheric $^{14}$C tend have a decay rate of $\sim$226 Bq/kg~\cite{sturm_carbon_2012}. For a sufficiently large organic liquid scintillator, this level of contamination can hide low energy signals that may be important for vetoing, reducing the overall neutron rejection power. Compounds made of methanol derived from petroleum, on the other hand, have much lower levels of $^{14}$C, because the carbon in them has spent long enough underground, shielded from cosmic activation, to allow the $^{14}$C to decay away and become heavily depleted.  Most of the $^{14}$C found in compounds from a petroleum origin is produced by underground nuclear reactions from $\alpha$ decays of uranium and thorium underground, followed by a ($\alpha$,n) then a (n,p) reactions on $^{14}$N. Borexino has reported a $^{14}$C contamination of 40 Bq in 100 tonnes of PC, for an overall $^{14}$C/$^{12}$C ratio of $10^{-18}$g/g~\cite{borexino_2013}, about six orders of magnitude lower than what would be expected from modern atmospheric carbon. Since most organic compounds can be made either way, it is important to check that the vendor uses a process with a mineral or ancient biogenic origin rather than a modern biogenic one in order to avoid this contamination.

\begin{figure}
 \centering
 \includegraphics[width=\linewidth]{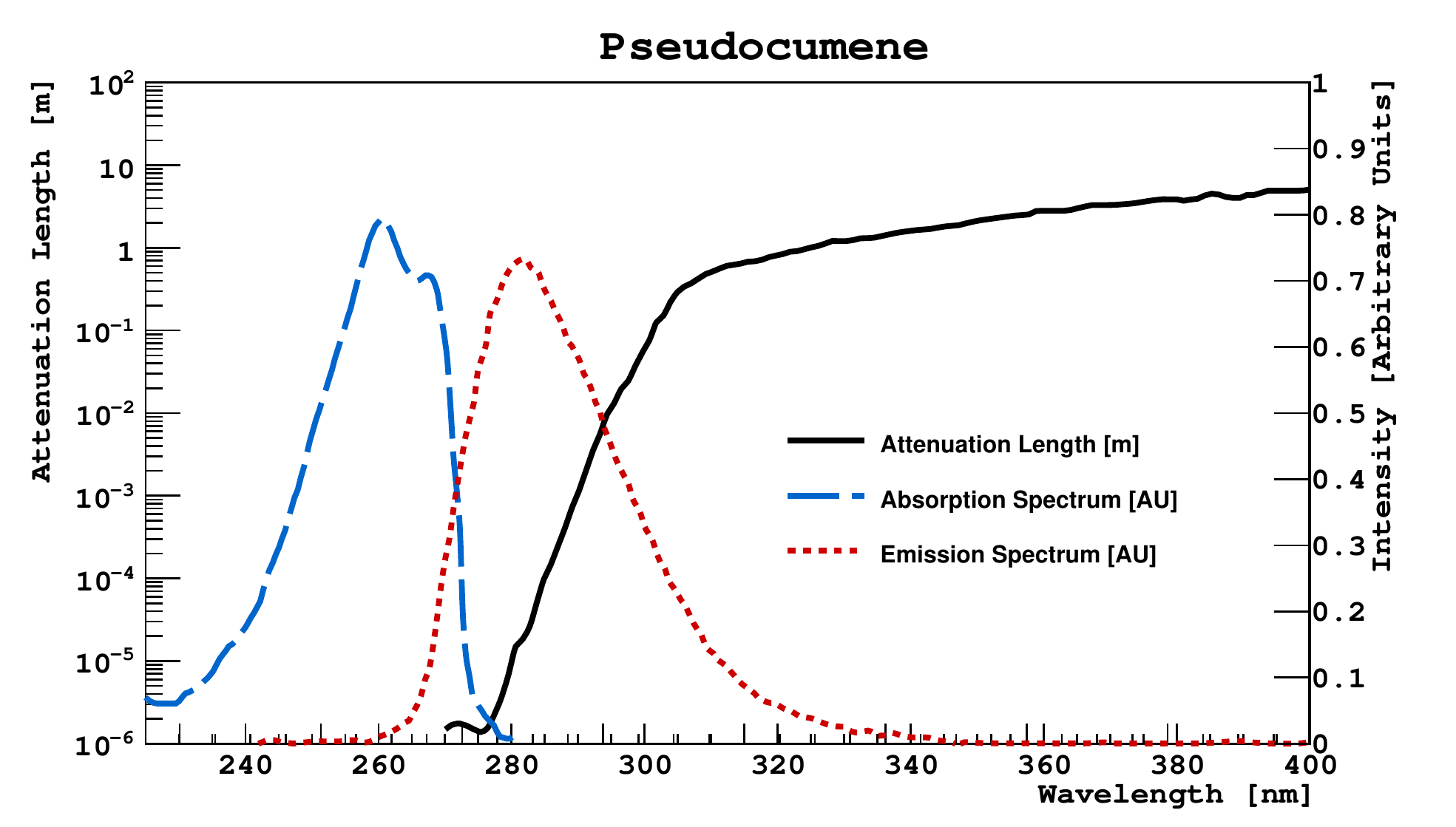}
 \caption{The attenuation length, emission, and absorption spectra of PC, measured by~\cite{borexjohnson}.}
 \label{pcspectra}
\end{figure}

\subsection{Pseudocumene}
Pseudocumene is an organic molecule with a single benzenoid ring that is known to scintillate in the UV region between 200--350 nm. 
The relationship between the absorption and emission spectra and the attenuation length of pure PC is shown in Fig.~\ref{pcspectra}.

\subsubsection{Oxygen Quenching:}
The presence of impurities in pseudocumene can result in a quenching effect that decreases the fluorescent light output. Oxygen is a particularly strong quenching agent, and even small amounts of oxygen contamination have a drastic effect~\cite{Borex1997}. Large amounts will turn PC yellow. Even when stored in an airtight container, it is possible for oxygen to enter the system by outgassing from the container materials. In order to rid PC of any oxygen that may have dissolved into it, one can distill PC under a vacuum, or sparge PC by bubbling nitrogen gas into it.

\subsubsection{Metal Interactions:}
\label{pcsssect}
A special case of oxygen quenching results from interactions between the PC and metal oxides in some materials. Some information regarding PC's compatibility with various metals can be found in~\cite{balseal}.

Effects of PC interacting with stainless steel have been observed to cause the attenuation length of light in PC to decrease over time~\cite{jayben}.
\begin{figure}
 \centering
 \includegraphics[width=.9\linewidth]{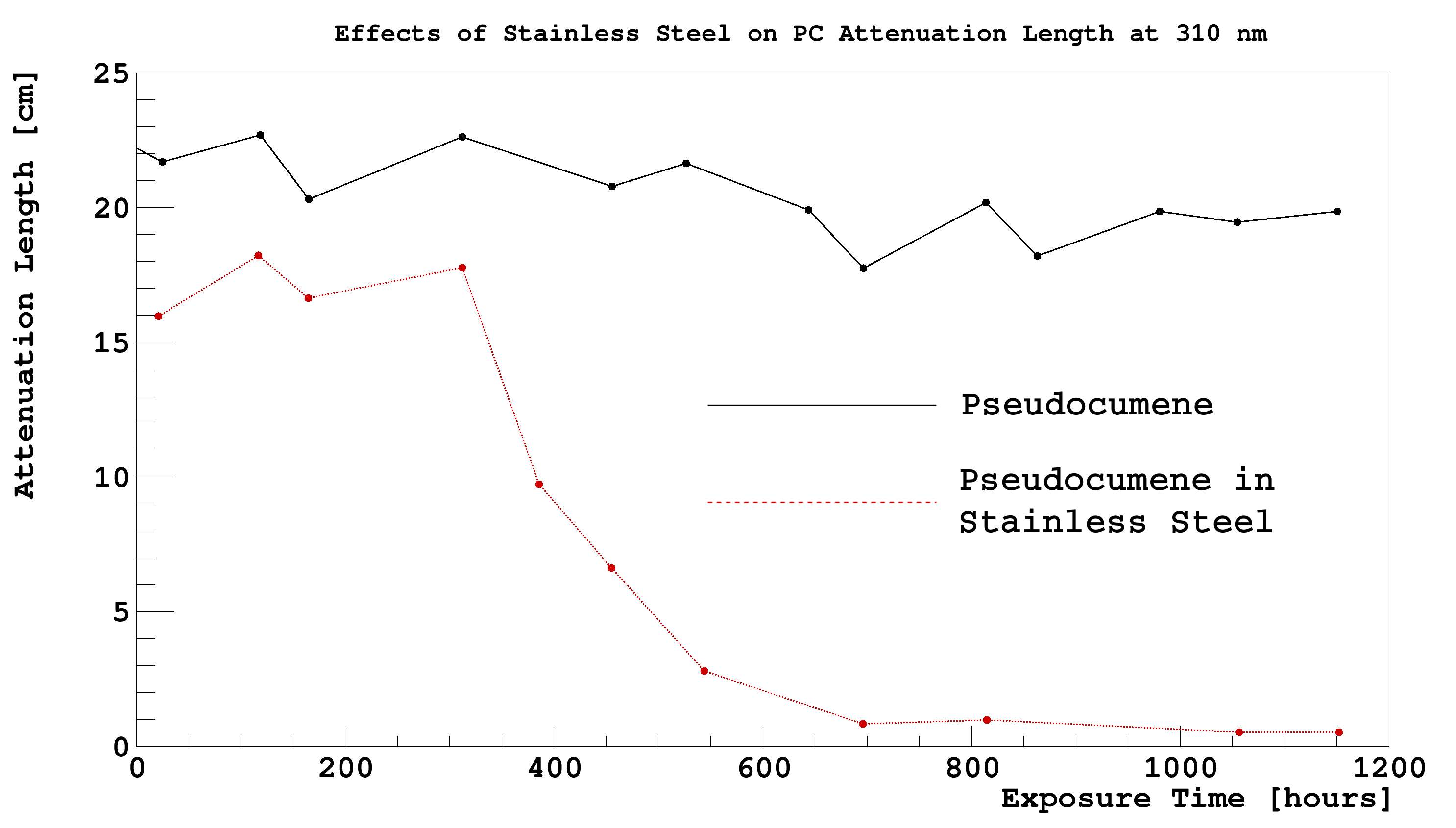}
 \caption{The degradation of the attenuation length of PC over time when exposed to a surface-to-volume ratio of 20 m$^{-1}$ of electropolished stainless steel compared to PC left unexposed. In order to speed up the interactions, the sample in stainless steel was heated by 50$^\circ$C for the first 300 hours and then by $100^\circ$C afterwards, from~\cite{jayben}.}
 \label{pcinss}
\end{figure}
Fig.~\ref{pcinss} shows the degradation of PC when exposed to a surface-to-volume ratio of 20 m$^{-1}$ of electropolished stainless steel for 24 days. After 400 hours, the attenuation length of 310 nm light drastically decreases.

Large vessels with lower surface-area-to-volume ratios are expected to show less degradation over time. For experiments that last a long time, this effect may still be significant and require either coating the stainless steel to prevent the reaction or repurifying the scintillation mixture multiple times during operation.

\subsection{Trimethyl Borate}
\subsubsection{Mixing with PC:}
Studies of adding TMB to PC + 10 g/L of 1-phenyl-e-mesityl-2-pyrazoline (PMP, a wavelength shifter similar to PPO) show that a solution containing equal parts PC and TMB will have a light yield approximately 85\% of pure PC~\cite{aldo_pctmb}. Additionally, since TMB is largely optically transparent to the light emitted by PC and PPO, adding TMB to the scintillator may increase the attenuation of light in the scintillator.
A mixture containing equal parts TMB and a primary solvent such as xylene or PC can still efficiently detect neutrons~\cite{birks}.

\subsubsection{Neutron Path Length:}
\begin{figure}
 \centering
 \includegraphics[width=\linewidth]{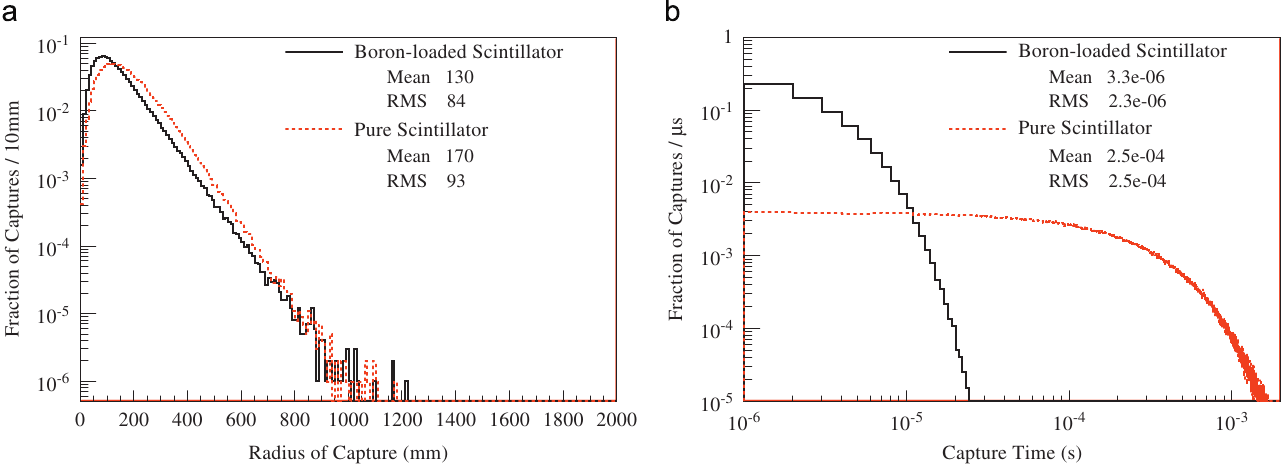}
 \caption{Simulated neutron capture radii and times for neutrons produced at the center of a spherical volume of an equal parts PC and TMB mixture, compared to the same in a mixture of pure PC, by~\cite{Wri2010}.}
 \label{tmbcapture}
\end{figure}

GEANT4 simulations  of radiogenic and cosmogenic neutrons (typically in the energy range of 1--10 MeV and 100 MeV, respectively) interacting with PC and TMB by~\cite{Wri2010} have yielded the neutron capture radius and time distributions shown in Fig.~\ref{tmbcapture}.

For these simulations, a 2 m radius spherical volume with a neutron source at the center was used. One of two scintillator mixtures used was pure PC, and the other was 50\% PC and 50\% TMB. The energies of the neutrons produced by the source were drawn from the measured and simulated energy distributions of cosmogenic neutrons, as well as radiogenic neutrons produced from the photodetectors. 

These simulations show that for the boron loaded scintillator, a mean neutron capture time of approximately 3.3 $\mu$s and a mean capture length of 13 cm can be expected. For a neutron veto with a radius of 2 m, the boron loaded scintillator can be expected to capture over 99.99\% of the neutrons that pass through it~\cite{Wri2010}.

\subsubsection{Water Contamination:}
TMB is highly hydroscopic, and will form boric acid when brought into contact with water through the reaction given in Equation \ref{boricacid}.
\begin{equation}
 (\mbox{CH}_3\mbox{O})_3\mbox{B}+3\mbox{H}_2\mbox{O}\rightarrow\mbox{H}_3\mbox{BO}_3+3\mbox{CH}_4\mbox{O}
 \label{boricacid}
\end{equation}
Boric acid is a fine white powder, and may serve as a contaminant in the scintillator, decreasing its attenuation length. It is therefore important that the TMB be distilled and contained in a moisture-free environment at all times. All surfaces that come in contact with TMB should also be made as dry as possible. 
\subsection{PPO}
The addition of a small concentration of a primary wavelength shifter such as PPO can greatly enhance the quantum yield of a liquid scintillator. By increasing the wavelength of the emitted light, PPO can also reduce self-absorption and increase the effectiveness of the reflector and photomultiplier tube (PMT) light detector (discussed in Section~\ref{pmtsect}).

Studies have been performed to measure the effects of PPO concentration on light yield for linear alkylbenzene~\cite{nemchenok}.

The light yield becomes roughly independent of PPO concentration for mass fractions above 0.3\%. In most of the studies presented here, we used a concentration of 3 g/L PPO, which corresponds to a 0.33\% mass fraction.

\begin{figure}
 \centering
 \includegraphics[width=\linewidth]{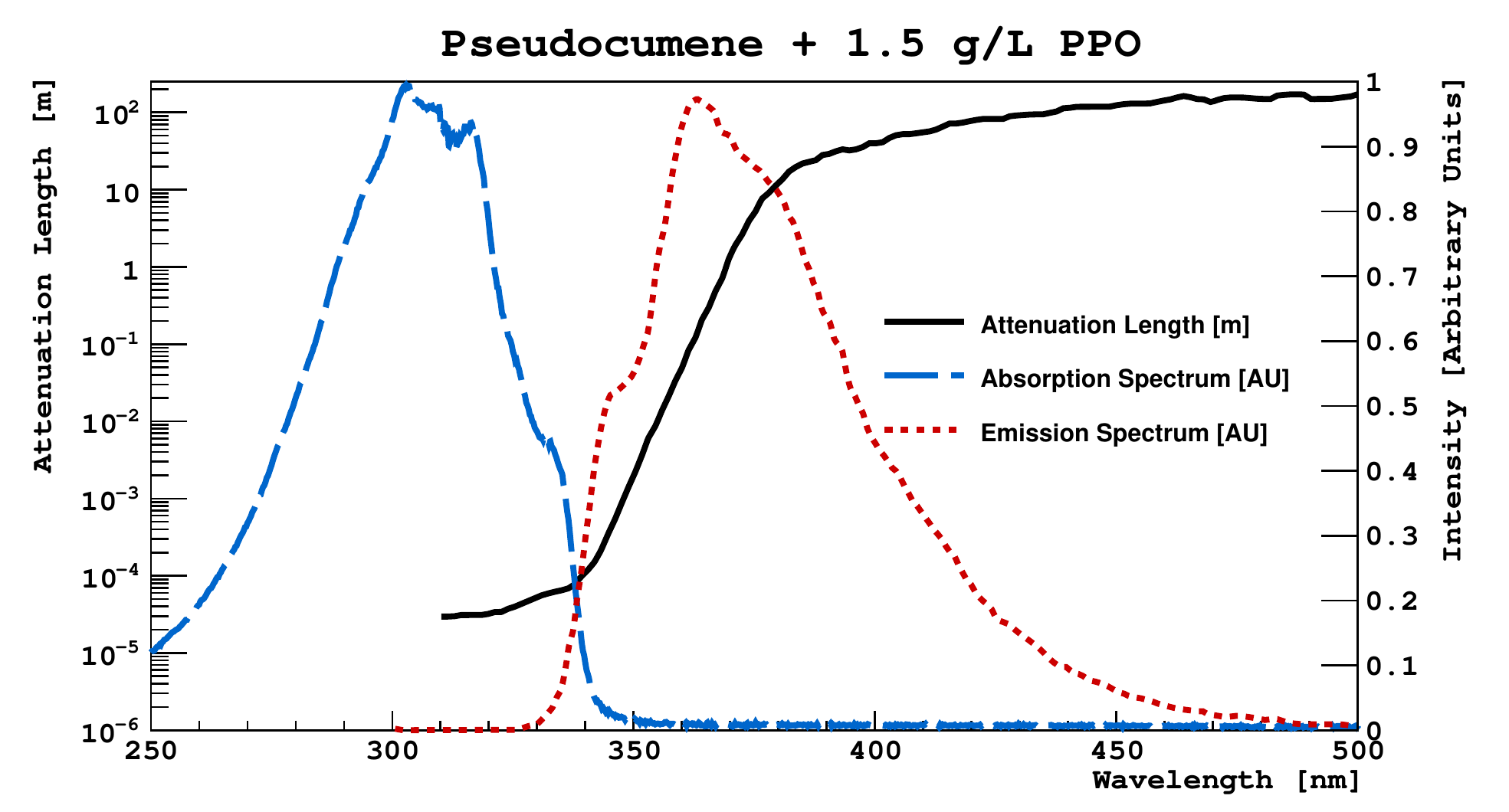}
 \caption{The attenuation length and emission spectrum of a mixture of PC +  1.5 g/L of PPO, measured by~\cite{borexjohnson}, and the absorption spectrum of PPO, measured by~\cite{Berlman}.}
 \label{ppospec}
\end{figure}
Fig.\ref{ppospec} shows the measured attenuation length, absorption, and emission spectra of 1.5 g/L of PPO dissolved in PC (the concentration used by Borexino). PPO drastically increases the attenuation length of light in the scintillator to approximately 10 m.

Additionally, measurements by Borexino in~\cite{Borex1997} show that increasing the PPO concentration in a liquid scintillator can decrease the scintillation time constant and decrease the ionization quenching of $\alpha$ depositions, which is particularly relevant for detecting the ground state products for neutron captures on $^{10}$B.

\subsection{POPOP and Bis-MSB}
In addition to a primary wavelength shifter, secondary wavelength shifters may be added to further push the emitted light to longer wavelengths, away from the absorption spectrum of PC and PPO. 

\begin{figure}
 \centering
 \includegraphics[width=\linewidth]{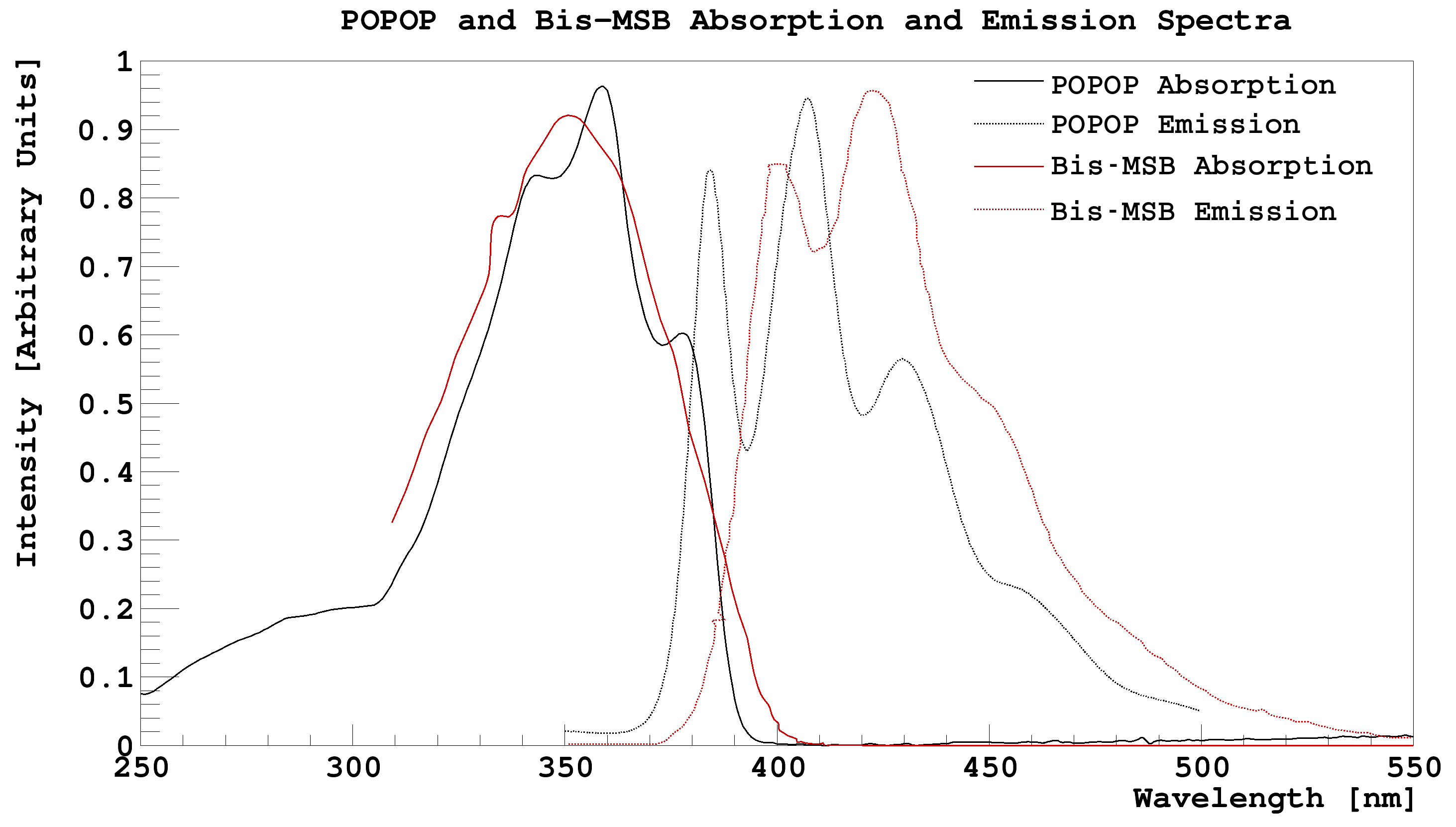}
 \caption{The absorption and emission spectra for POPOP dissolved in cyclohexane~\cite{Berlman}, and bis-MSB dissolved in PC~\cite{borexjohnson}.}
 \label{wlsspec}
\end{figure}

The relative light yields stop increasing with concentration at approximately 0.0025\% mass fraction for POPOP and 0.001\% mass fraction for bis-MSB~\cite{nemchenok}. These roughly correspond to concentrations of approximately 0.25 g/L of POPOP and 0.15 g/L of bis-MSB.  Fig.\ref{wlsspec} shows the absorption and emission spectra of POPOP and bis-MSB. 

\section{Reflector}

Because high photomultiplier coverage would be expensive for most detectors, the veto concept relies on highly reflective surfaces to maintain a high light-collection efficiency.  

It is important that the reflector used for the veto be chemically compatible with the scintillator and that it can retain a high reflectivity over an extended period of submersion. It must also be cheap enough to effectively cover the large surface area of the veto.

\subsection{The Spectrophotometer}
Reflectance measurements of potential reflectors were made on a Perkin Elmer Lambda-650 spectrophotometer, which measures reflectance in the range of 200--860 nm at an 8$^\circ$ angle from the reflector surface. While these measurements were only taken at this angle of incidence, the angular dependence of many common reflectors has been measured by others~\cite{Jan2012}. Reflectance measurements were normalized to that of a Spectralon plug, which is nearly 100\% reflective~\cite{LabSphereTech}.  The spectrophotometer is very stable over time if the plug is kept clean in a dark location.  Re-measuring standard samples several months apart show variations in measured reflectance around 0.1\%. 

\subsection{Reflector Candidates}
\label{reflcandsect}
Several potential reflectors were measured in the spectrophotometer.  Candidates were chosen based on their reflectance when dry.  These measurements were conducted on several types of Tyvek (a void-based, paper-like reflector made of long  polyethylene chains), Duraflect (a void based reflector made from a compacted PTFE powder), Crystal Wrap (a thin PTFE-based film with a high void fraction), high purity aluminum foil, 3M Foil, and Lumirror (a PET-based multi-layer film used in LCD screens).

Void-based reflectors work by exploiting the different indices of refraction between the voids and the material themselves. These different indices of refraction cause light to refract and diffusely reflect back. Many of the void-based substances were eliminated from consideration because of a considerable loss in reflectance when soaked in scintillator. This is likely due to the air in the voids being displaced by scintillator, which has a much closer index of refraction to the materials of the reflectors. This causes the reflector to become translucent.
3M Foil, which is highly reflective, had to be eliminated due to its reactivity with trimethyl borate.  The aluminum was dropped from consideration because its reflectance was not competitive with the other two remaining reflectors, even when they are wet.  

\begin{figure}
\centering
\includegraphics[width=.91\linewidth]{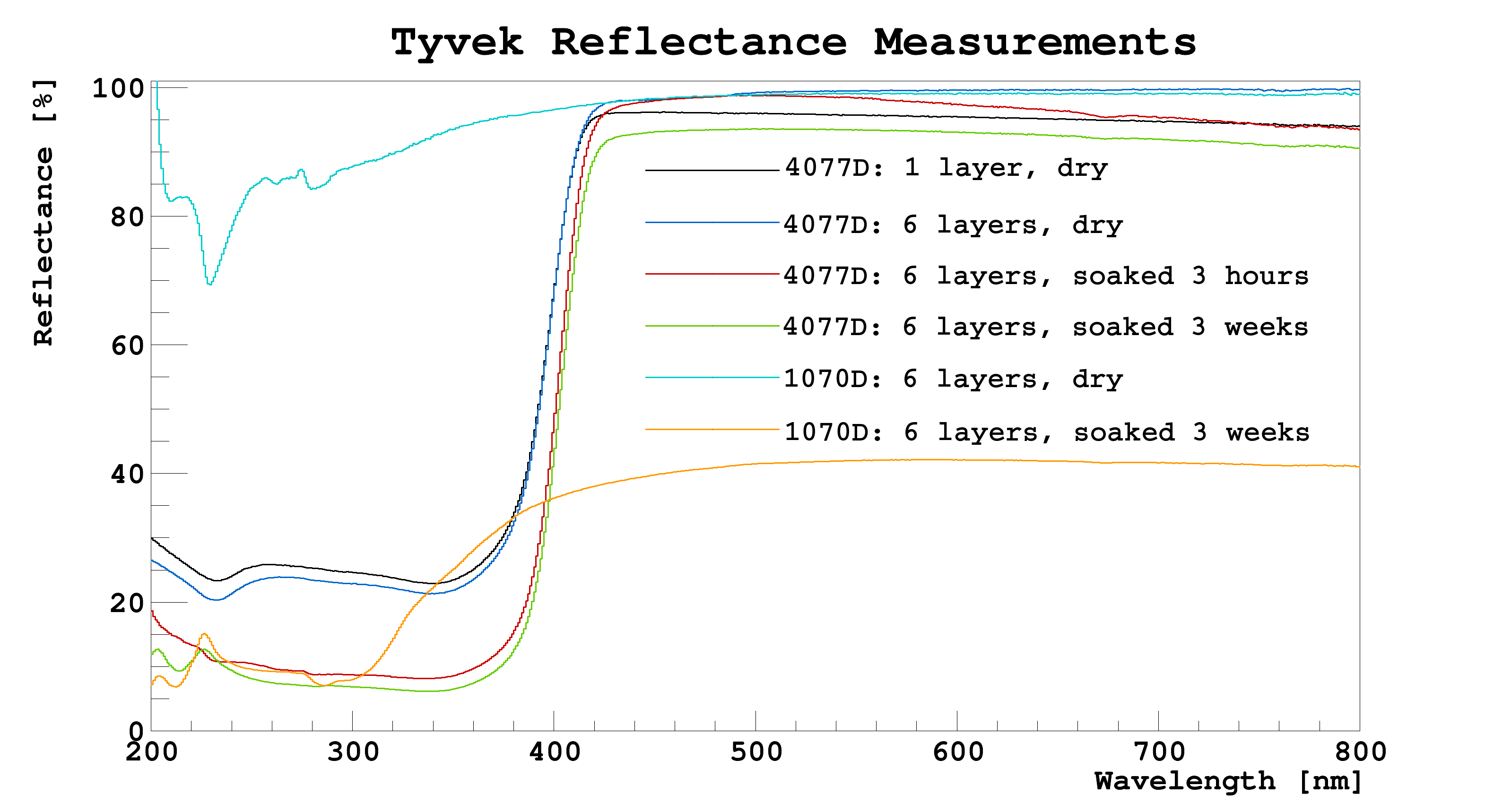}
\caption{Reflectance measurements of Tyvek reflectors. The black curve shows the reflectance of Tyvek 4077D with one layer, while blue shows the same with 6 layers. Additional curves compare two different types of Tyvek after soaking in a PC+TMB cocktail for various lengths of time: (red) 4077D after 3 hours, (orange) 4077D after 3 weeks, (green) 1070D before soaking, and (cyan) 1070D after 3 weeks. (Not shown) Tyvek 1082D exhibited very similar behavior to 1070D.}
\label{tyvek}
\end{figure}

\begin{figure}
\centering

\includegraphics[width=.91\linewidth]{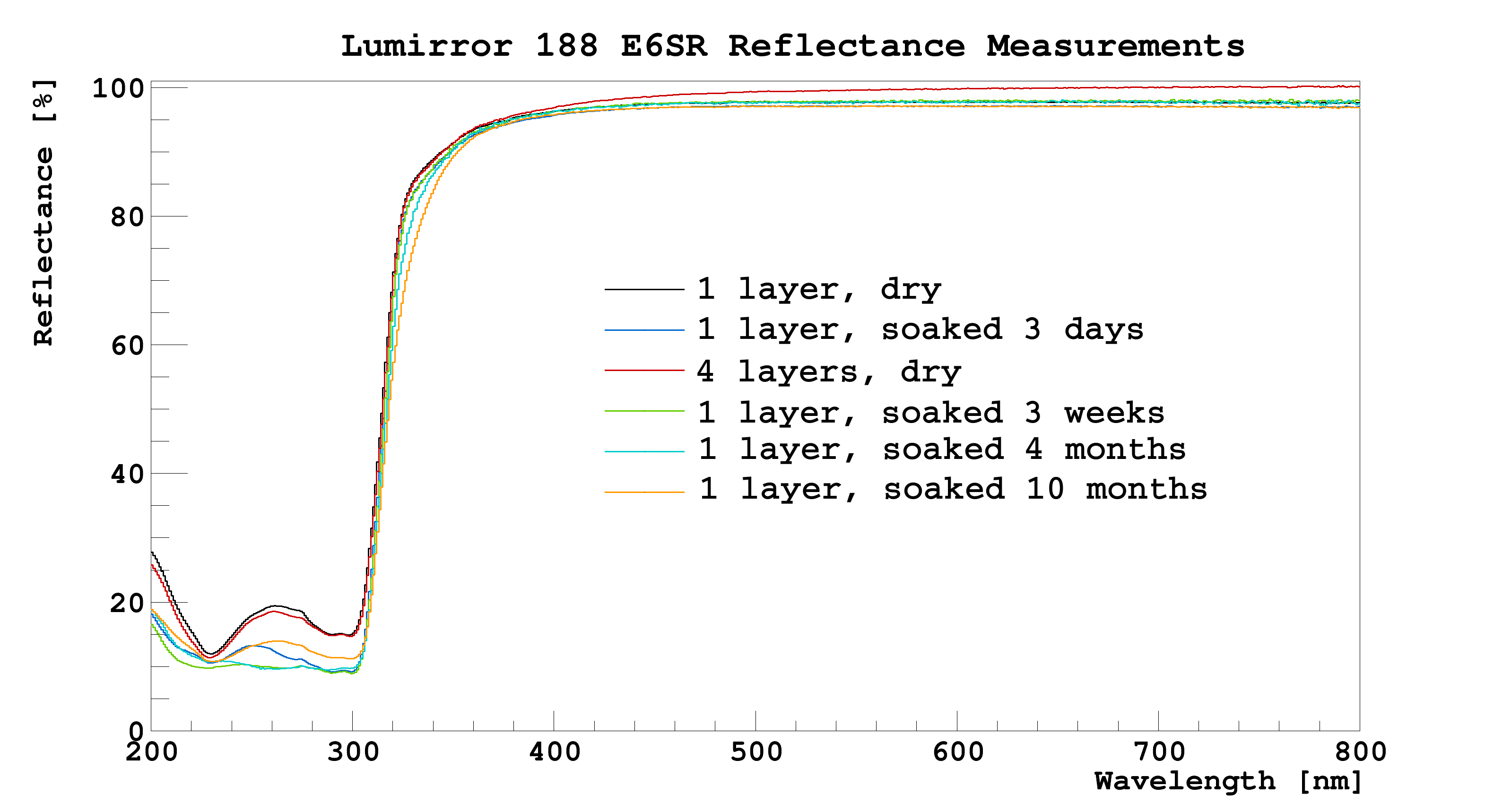}
\caption{Reflectance measurements of Lumirror 188 E6SR, after soaking in a PC+TMB cocktail for varying lengths of time: (black) before soaking, (blue) after 3 days, (orange) after 3 weeks, (green) after 4 months, and (cyan) after 10 months. The red curve shows the improvement of a dry sample while using 4 layers. The bump around 240 nm is due to the specular component of the reflectivity.}
\label{lumirror}
\end{figure}

The remaining reflector candidates were Tyvek 4077D and Lumirror 188 E6SR, whose reflectances are shown in Figs.~\ref{tyvek} and~\ref{lumirror}, respectively. Table~\ref{refltable} summarizes the measurements shown in these figures. Tyvek 4077D has a titanium-oxide component that keeps scintillator from filling the voids, but introduces a severe cutoff in reflectance at 400 nm. This can be seen in Fig.~\ref{tyvek}, where the reflectance of Tyvek 4077D is compared to that of Tyvek 1070D, which does not have the TiO. The dry Tyvek 1070D has a higher reflectance than the Tyvek 4077D, and does not have such a steep high energy cutoff. However, as can be seen from this figure, Tyvek 1070D loses most of its reflectance when soaked in scintillator, while Tyvek 4077 loses considerably less (though the losses are still around 6--7\%).

\begin{figure}
\centering
\includegraphics[width=.4\textwidth]{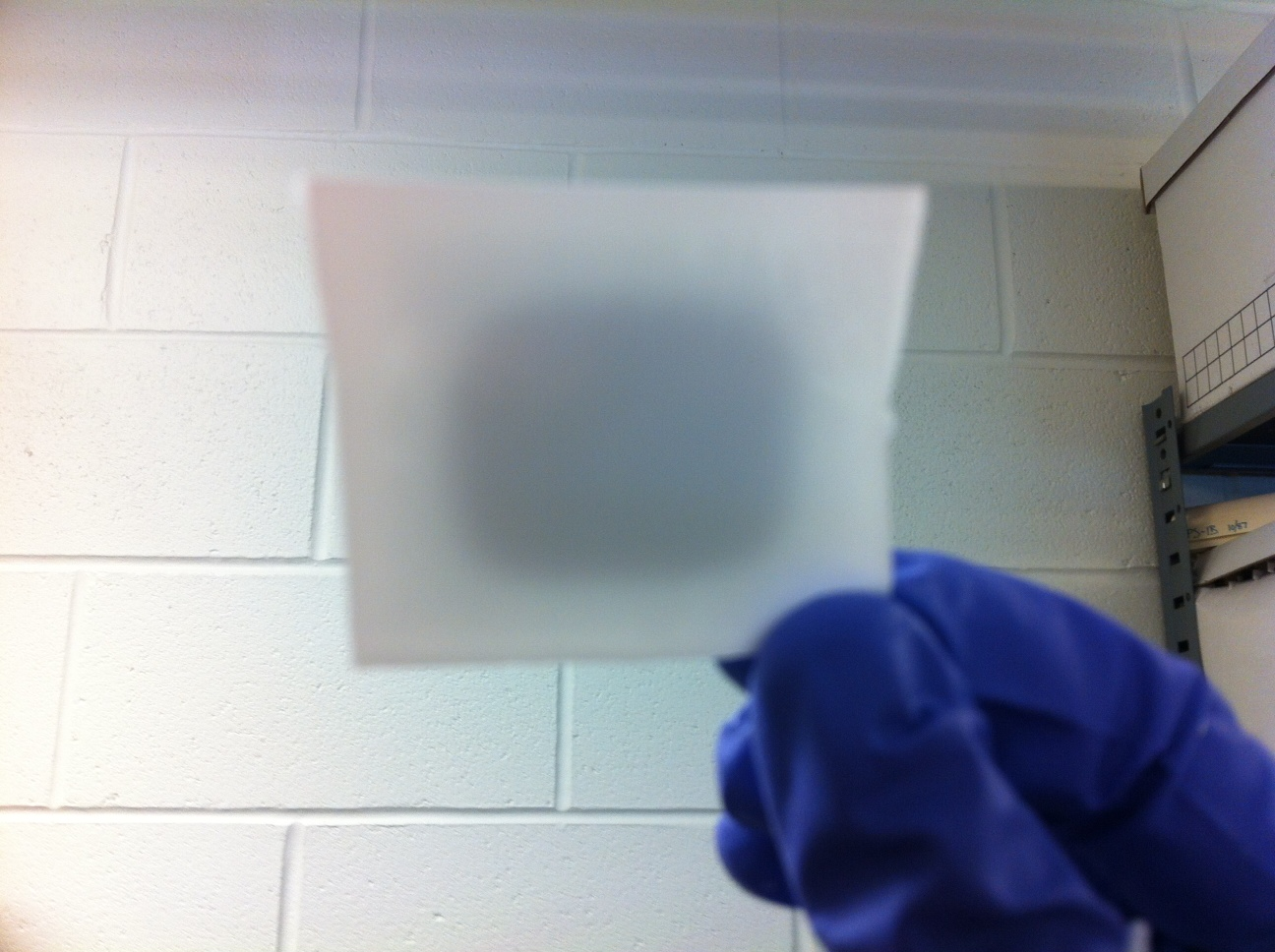}
\caption{A small sample of Lumirror 188 E6SR that has been soaked in scintillator for nine months.  Some degradation of the reflectance at the edges has occurred due to the creeping of scintillator through the sides.  The reflectance of the central part of this sample remains unaffected, however (see Fig.\ref{lumirror}).}
\label{edges}
\end{figure}

The Lumirror has a lower cutoff wavelength (around 325 nm) and a very high reflectance when dry (98\%). Because of its multi-layer-based reflectance, it is very minimally affected by the scintillator.  Though the bulk of the material is protected from the scintillator, some creeping of scintillator into the sides of the material has been observed, which lowers the reflectance significantly near the edges.  This effect is very slow however, progressing about 1 cm over the course of nine months (see Fig.\ref{edges}). Measurements of the reflectance of the Lumirror after this degradation show that the edges drop to a peak reflectance of $\sim$83\% at 380 nm, while the bulk remains unaffected. To circumvent this loss in reflectance, a larger detector can overlap layers of Lumirror by a few centimeters, so the degraded edges will be covered by the still highly reflective bulk of another layer. Lumirror's reflectance under various conditions is shown in Fig.\ref{lumirror}.  Comparing the black line (dry, with a maximum reflectance around 98\%) with the blue (soaking in PC+TMB for 3 days), purple (3 weeks), green (4 months), and cyan (10 months) show the reflectance dropping down by about 0.7\% after 3 days, going back to the original value, and then returning to the lower value after 10 months. This behavior is likely due unintended variation in the procedure between these measurements, leaving two possible interpretations. Either the reflectance has remained stable after being soaked in scintillator, or it quickly decreased and stabilized at a slightly lower reflectance. This may be due the scintillator penetrating a small distance into the Lumirror, and changing its reflectance due to the scintillator's different index of refraction. However, it is worth commenting that even in the case where the scintillator does decrease the reflectance, the loss is less than a percent, leaving Lumirror as a strong candidate.

It has also been shown that Lumirror fluoresces with primary absorption between 320 and 420 nm and emits around 440 nm~\cite{Jan2012}.  
This fluorescence acts as an additional wavelength shifter and may increase our light collection.  The efficiency of this fluorescence is unreported. The reflectance of both the Tyvek and the Lumirror can be improved by adding additional layers of the material.

\begin{table}
\centering
\caption{Summary of the Tyvek and Lumirror reflectances shown in Figs.~\ref{tyvek} and~\ref{lumirror}. $R_{250}$ and $R_{500}$ are the reflectances measured at 250 nm and 500 nm, respectively. $\lambda_{50\%}$ and $\lambda_{90\%}$ are the wavelengths at which each reflector had 50\% or 90\% reflectance, respectively. Notably, the six layers of Tyvek 1070D was above 50\% for the entire range of wavelengths measured while dry, and the second time it passes the 90\% reflectance mark is reported. While it is dry, this sample is below 50\% reflectance for the entire range of wavelengths measured.}
\begin{tabular}{l|cccc}\hline
Description & $R_{250}$ [\%] & $R_{500}$ [\%] & $\lambda_{50\%}$ [nm] & $\lambda_{90\%}$ [nm]\\\hline\hline
\multicolumn{5}{c}{Tyvek 4077D}\\\hline
1 layer, dry & 25.5 & 96.0 & 392 & 411 \\
6 layers, dry & 23.1 & 99.2 & 392 & 411 \\
6 layers, soaked 3 hours & 10.4 & 98.7 & 401 & 416 \\
6 layers, soaked 3 weeks & 8.1 & 93.6 & 403 & 422 \\\hline
\multicolumn{5}{c}{Tyvek 1070D}\\\hline
6 layers, dry & 85.2 & 93.2 & ---  & 335 \\
6 layers, soaked 3 weeks & 9.5 & 41.5 & --- & --- \\\hline
\multicolumn{5}{c}{Lumirror 188 E6SR}\\\hline
1 layer, dry & 18.0 & 97.8 & 315 & 344 \\
4 layers, dry & 17.3 & 99.4 & 315 & 345\\
1 layer, soaked 3 days & 13.2 & 97.1 & 316 & 349 \\
1 layer, soaked 3 weeks & 10.1 & 97.8 & 316 & 348\\
1 layer, soaked 4 months & 10.0 & 97.6 & 317 & 349\\
1 layer, soaked 10 months & 13.2 & 97.1 & 318 & 352\\\hline
\end{tabular}
\label{refltable}
\end{table}

After conducting tests that showed the stability of the reflectance of Lumirror and its lack of reactivity with the scintillator, we decided to use Lumirror 188 E6SR (a 188-micron thick Lumirror film) as our reflector for our prototype neutron veto (see Section~\ref{belljarprep}).

After settling on Lumirror for the reflector, it was important to check that its natural radioactive contamination would not produce too high of a background rate. Samples of Lumirror were sent to Laboratori Nazionali del Gran Sasso where they were screened for radioactivity using the GeMPI germanium crystal detector described in~\cite{maneschg_measurements_2008}. The results from these measurements are given in Table~\ref{lumirrorScreening}. The DarkSide-50 neutron veto detector, with a diameter of 4 m, has $\sim$7 kg of Lumirror on the inner surface. The radioactivity of the samples was deemed low enough to make Lumirror a good choice of reflector for a large neutron veto. 

\begin{table}[tp]
 \centering
 \begin{tabular}{llc}\hline
  & Nuclide & Concentration [mBq/kg] \\\hline\hline
  $^{232}$Th decay chain  & $^{228}$Ra & 72$\pm$6 \\
  & $^{228}$Th & 31$\pm$3 \\\hline
  $^{238}$U decay chain & $^{226}$Ra & 657$\pm$26 \\
  & $^{234}$Th & $<$330\\
  & $^{234m}$Pa & 150\\\hline
  $^{235}$U decay chain & $^{235}$U & 9$\pm$4\\\hline
  & $^{40}$K & 430$\pm$50\\\hline
  & $^{137}$Cs & 8$\pm$1 \\\hline
  & $^{60}$Co & $<$1.8\\\hline
 \end{tabular}
 \caption{Radionuclide measurements made of a 1.7 kg sample of Lumirror E6SR (188 $\mu$m thick). Nuclides that are part of the same decay chain are grouped together~\cite{matthias_counting}.}
 \label{lumirrorScreening}
\end{table}

\section{Photomultiplier Tube}
\label{pmtsect}
Light yield measurements of each setup were taken using a 3'' diameter Hamamatsu R11065 photomultiplier tube, with a 6.5 cm $\phi$ flat, bialkali photocathode. 

\begin{figure}
 \centering
 \includegraphics[width=\linewidth]{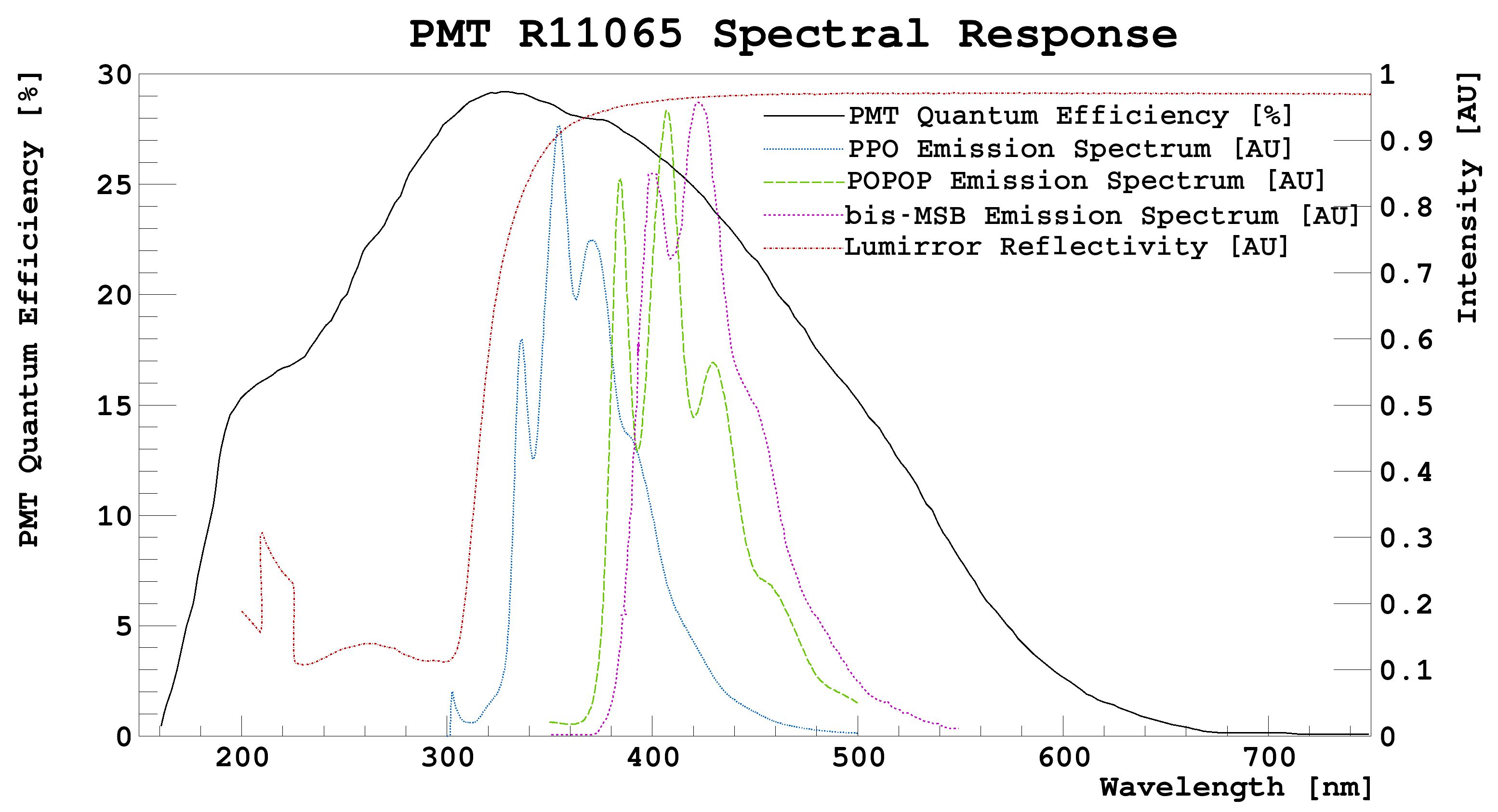}
 \caption{Plot of a typical quantum efficiency distribution for an R11065 PMT, provided by Hamamatsu~\cite{HamCat}, shown alongside the reflectance spectrum for Lumirror and the emission spectra of PPO~\cite{Berlman}, POPOP~\cite{Berlman}, and bis-MSB~\cite{bismsbref}, all scaled to arbitrary units.}
 \label{qeplot}
\end{figure}

The quantum efficiency of a typical R11065 PMT as quoted by Hamamatsu is shown in Fig.~\ref{qeplot}. As shown in the figure, the peak quantum efficiency of just under 30\% occurs at approximately 320 nm. The particular PMT used for these studies was measured to have a peak quantum efficiency of 31\% from 320--340 nm. The PMT continues to have a high quantum efficiency above 25\% around the peaks of all three wavelength shifters used in these studies, spanning the wavelength range of 340--420 nm. Additionally, the Lumirror reflectance flattens out at around 98\% near the peak quantum efficiency of the PMT. This agreement makes Lumirror and PPO with or without the addition of bis-MSB and POPOP good matches for optimizing the light yield with this type of PMT.

\section{Experimental Apparatus}

\subsection{The Bell Jar}
\label{bjsect}
\begin{figure}
\centering
\includegraphics[width=.45\textwidth]{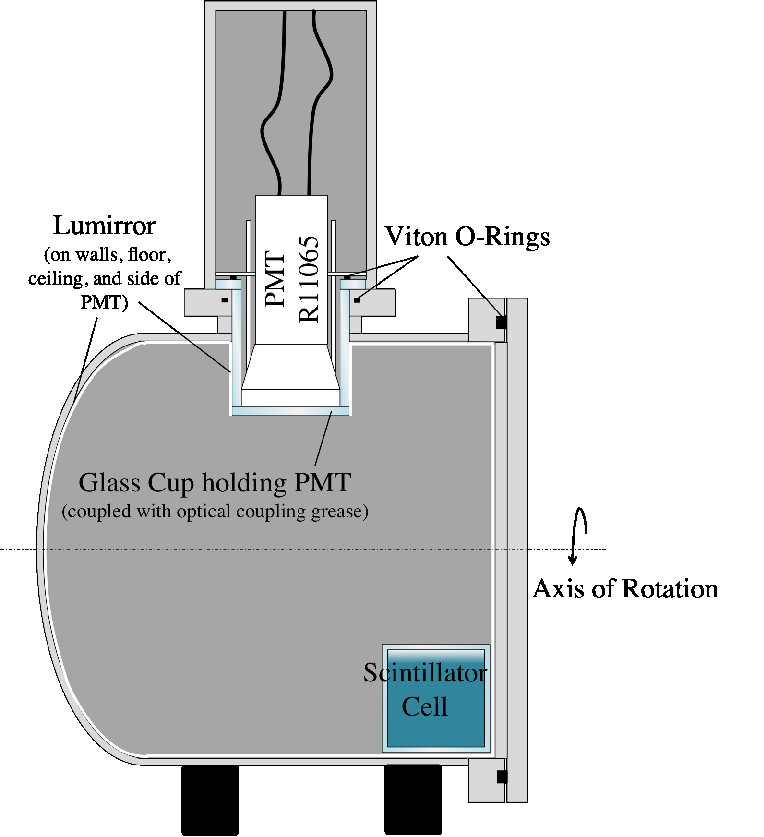}
\includegraphics[width=.45\textwidth]{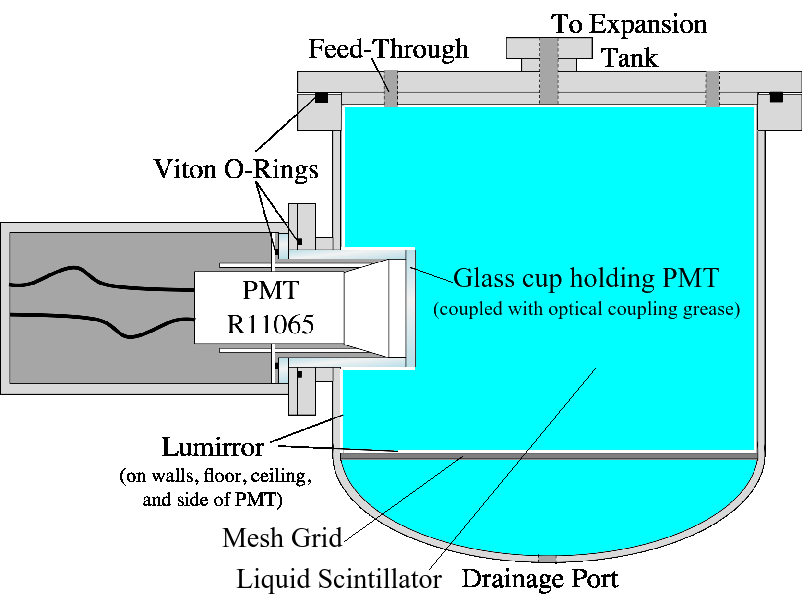}
\caption{Cross section of bell-jar setup for each experiment.  Our cell setup (left) shows a cell filled with scintillator in the otherwise dry bell jar.  The filling setup (right) shows the bell jar, which has been modified to be filled with scintillator.}
\label{belljar}
\end{figure}

A bell jar from an evaporator was adapted to create a closed detector as a prototype of the proposed neutron veto for DarkSide-50.  The bell jar is a 30.5 cm $\phi$ cylinder with a rounded section on top for a total height of 30.5 cm.  It is sealed shut with a lid that is clamped over a Viton o-ring.  A conflat port on the side allows for the insertion of a 3" R11065 Hamamatsu PMT.  This PMT is separated from the scintillator by a glass jar as shown in Fig.~\ref{belljar}.  Along the walls, ceiling, and floor of the bell jar, as well as the side of the jar housing the PMT, our reflector is wrapped and fixed to the surfaces by screws welded onto the bell jar and teflon nuts.  

\subsection{Modes of Operation}
The bell jar was used in two different modes in this experiment.  It was first tested dry with an isolated scintillator cell placed on the floor inside.  This setup was used as a preliminary test to determine the optimal scintillator cocktail and reflector to be used in the veto.  The cells were comprised of fused silica cylinders (7.6 cm in diameter, 7.6 cm tall), and were filled with the different scintillator cocktails.  In this configuration, the reflector lined the curved part of the bell jar.

The second series of tests were conducted with the bell jar filled with scintillator.  Several modifications to the initial setup were made.  A flat mesh grid was used to make the lining of the active volume with reflector easier.  Fittings were added to the bell jar for filling and emptying, as well as for an expansion tank which allows for fluctuations in the volume of the scintillator as a result of thermal expansion. 

\subsection{Electronics}
\begin{figure}[tb]
 \centering
 \includegraphics[width=\linewidth]{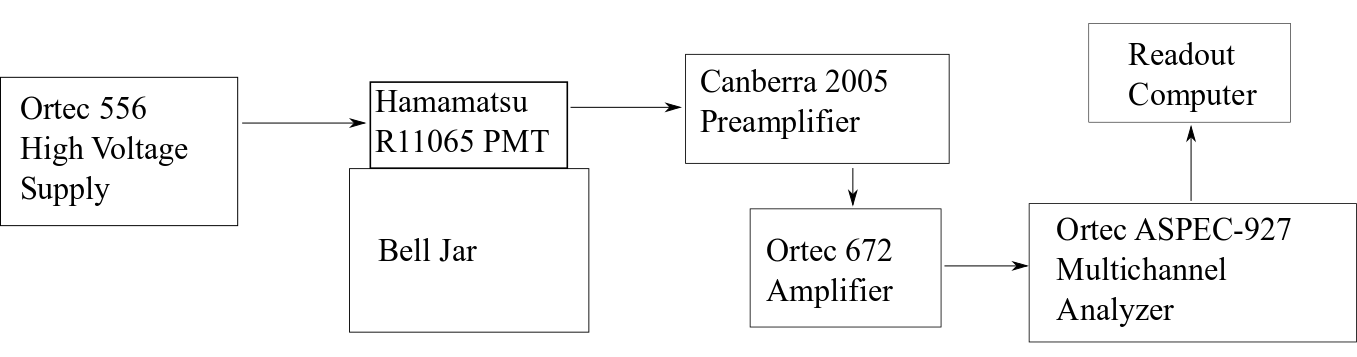}
 \caption{The data acquisition electronics setup for the bell jar measurements.}
 \label{elecdiagram}
\end{figure}

\begin{figure}[tb]
 \centering
 \includegraphics[width=0.9\linewidth]{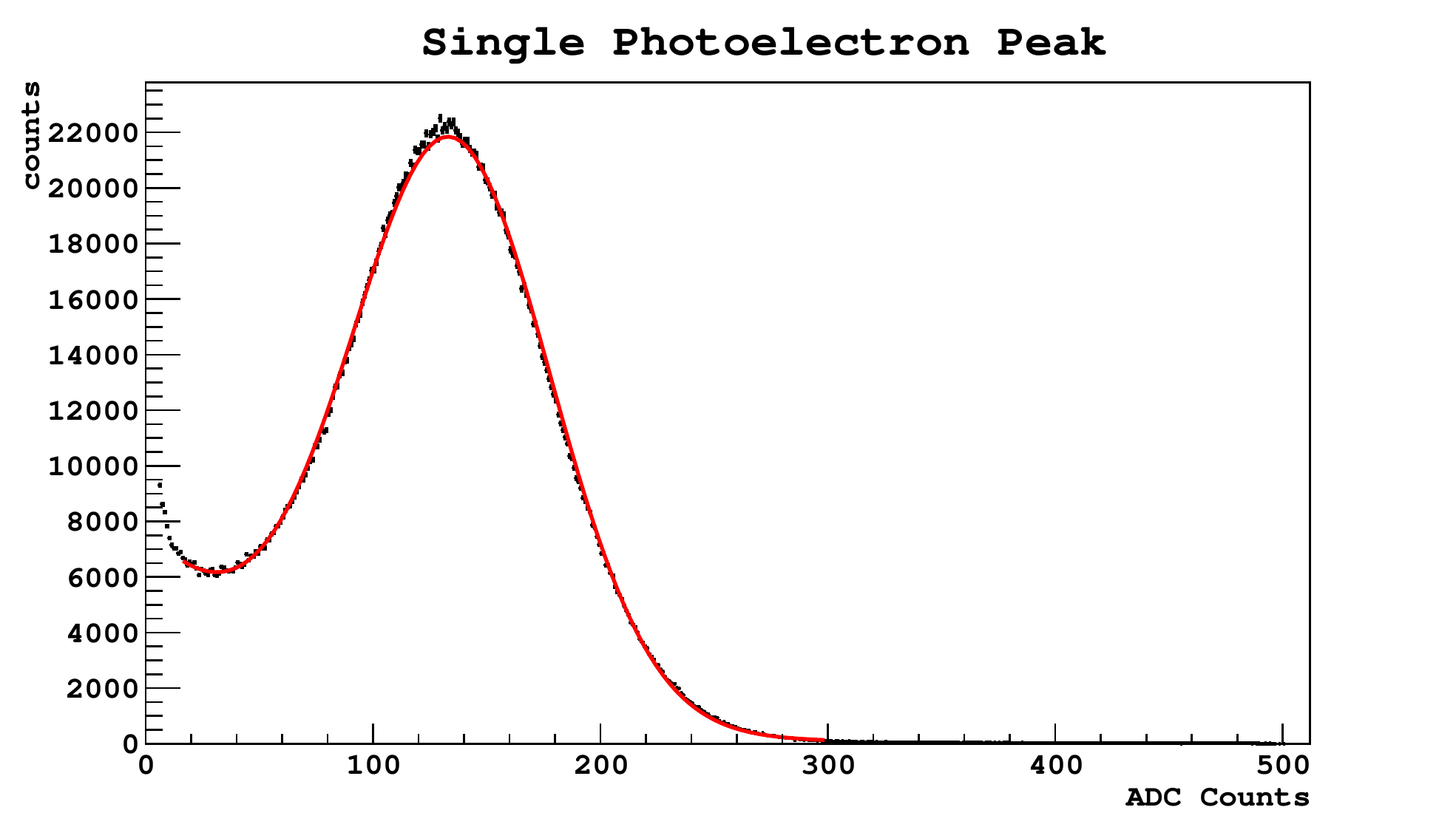}
 \caption{A sample single photoelectron spectrum measured with this setup (black). A Gaussian plus an exponential was fit to this spectrum to extract the mean number of ADC counts in a single photoelectron (red).}
 \label{spedist}
\end{figure}

The PMT was powered with an Ortec 556 high voltage power supply operated at positive high voltage. The signal from the PMT went through a Canberra 2005 preamplifier to an Ortec 672 amplifier operated with a 0.5 $\mu$s shaping time. Single photoelectron peaks were measured with a gain of 1.5$\times$1000. $\gamma$ spectra were measured at gains 
100 times smaller.  The unipolar triangular output of the amplifier was passed to an Ortec ASPEC-927 multichannel analyzer, which then transmitted data to the computer, where it was recorded using Ortec's Maestro multichannel analyzer software. Fig.\ref{elecdiagram} shows a block diagram of the electronics setup used for these measurements.

Fig.\ref{spedist} shows a sample single photoelectron distribution measured with this setup. A Gaussian plus an exponential tail at low counts was used to determine the number of ADC counts per photoelectron. This number was used for future measurements when converting the number of ADC counts observed in each energy bin into a number of photoelectrons.

\section{Light Yield Tests with Small Scintillator Cells}
\label{smallcellsect}
\begin{figure}
 \centering
 \includegraphics[width = .3\linewidth]{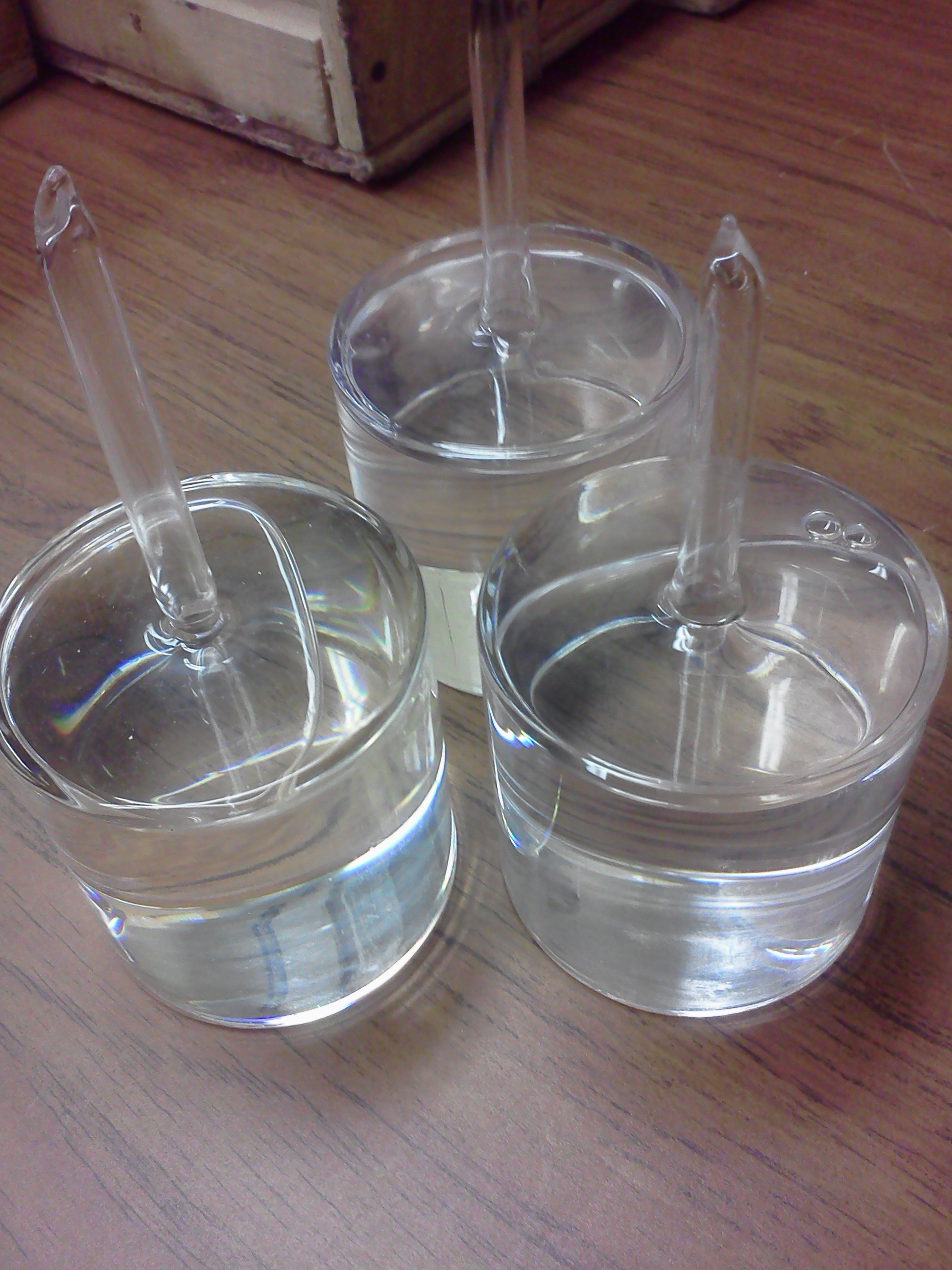}
 \caption{Three of the small test cells used for determining the optimal scintillator mixture.}
 \label{smallcells}
\end{figure}

In order to determine the optimal scintillator mixture, a series of 7.62 cm $\phi\times$7.62 cm high clear fused quartz cylinders were constructed, as shown in the first configuration in Fig.\ref{smallcells}. 
\subsection{Cell Preparation}
\begin{figure}
 \centering
 \includegraphics[width=.65\linewidth]{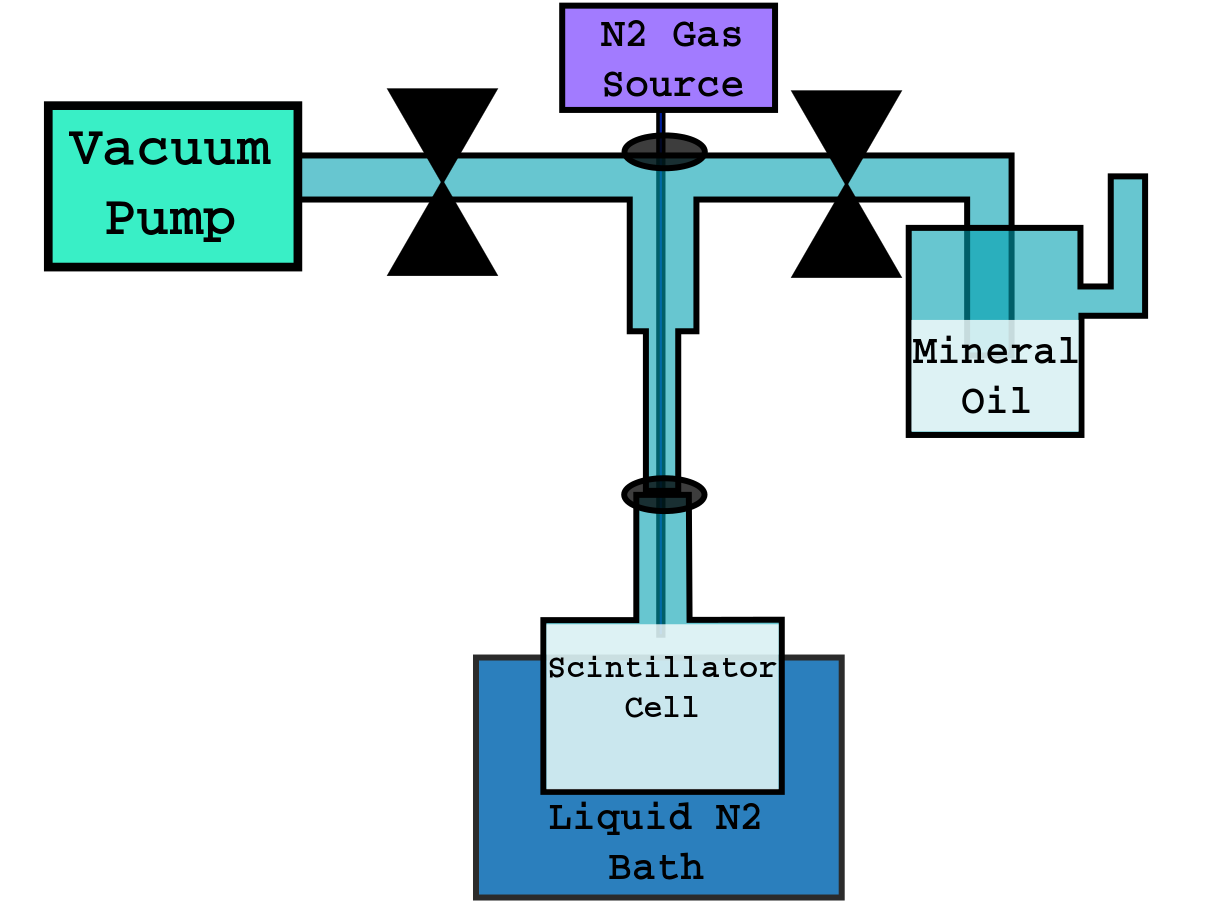}
 \caption{The stand used to sparge and seal the small test cells. Glass piping connects the cell to a vacuum pump and a container of mineral oil that allows air to flow out of the piping but not in. A metal rod connected to a nitrogen source can also be inserted into the top of the piping through a septa pad. The rod can be lowered into the scintillator to sparge it.  The cell is positioned to be lowered into a nitrogen path. Valves can be used to control the airflow in the pipes to the vacuum pump and mineral oil.}
 \label{fillcell}
\end{figure}

Before the cells were filled, the PC and TMB were separately distilled into flasks. After distillation, each flask was immediately capped and transferred into a glove box with a constant nitrogen flow maintaining a relative humidity between 3\% and 6\%
. The scintillator components were mixed together and poured into the cell.  The top of the neck was capped off, and the cell was taken out of the glove box and quickly moved to a nitrogen-filled stand shown in Fig.\ref{fillcell}.
 After attaching the cell to the apparatus, we then sparged the cell by bubbling nitrogen into it, allowing gas to escape the system through the mineral oil without letting air flow in. 

Once the scintillator had been sparged, it was submerged into a tank of liquid nitrogen. After the scintillator froze, the entire system was brought to a vacuum while the cell was sealed shut with a torch. 

\subsection{Measurements}
The bell jar was turned on its side and allowed to rotate, so that different cell positions could be tested (Fig.\ref{belljar}). The cell was placed at the flat base of the bell jar. The $\gamma$ spectrum was measured for each cell, followed by a measurement of the single photoelectron peak of the cell. The location of the Compton edge of each cell and the single photoelectron peak were used to determine the light yield in photoelectrons detected per keV deposited in the cell.

Light yield measurements taken with the cell across from the PMT for different scintillator cocktails and the two reflector candidates are shown in Table~\ref{smallresults}.  Light yield varied less than 5\% as a function of position.
 \newlength{\longline}
 \settowidth{\longline}{PC+TMB+PPO (1.5 g/L) +}
 \begin{table}
 \centering
 \begin{tabular}{|l|c|c|}
 \hline
  \multicolumn{3}{|l|}{\textbf{Light Yields (p.e./keVee)}}\\\hline
  Cocktail & Tyvek & Lumirror\\\hline\hline
  PC+PPO (1.5 g/L) & 0.115$\pm0.008$ & 0.472$\pm$0.019\\\hline
  PC+TMB+PMP (5 g/L) & 0.403$\pm0.015$ & ---\\\hline
  \parbox[t]{\longline}{PC+TMB+PPO (1.5 g/L) +\\ POPOP (25 mg/L)} & 0.286$\pm0.014$ & 0.301$\pm0.011$\\\hline
  \parbox[t]{\longline}{PC+TMB+PPO (1.5 g/L) +\\ bis-MSB (25 mg/L)} & 0.267$\pm0.015$ & 0.336$\pm0.013$\\\hline
  \parbox[t]{\longline}{PC+TMB+PPO (3 g/L) +\\ POPOP (25 mg/L)} & --- & 0.408$\pm0.011$\\\hline 
  \parbox[t]{\longline}{PC+TMB+PPO (3 g/L) +\\ bis-MSB (15 mg/L)} & --- & 0.389$\pm0.016$\\\hline
 \end{tabular} 
 \caption{Measured light yields of the bell jar in the scintillator cell configuration with Tyvek or Lumirror as the reflector in photoelectrons/keV.  For all measurements the bell jar was filled with air and the scintillator cell was placed opposite the PMT. Measurements were taken with a $^{54}$Mn $\gamma$ source.}
 \label{smallresults}
 \end{table}
Errors in light yield measurements come from spread in the Compton edge due to the finite resolution of the system.

These studies showed that Lumirror is much more effective than Tyvek 4077D both wet and dry, and that 3 g/L of PPO results in higher light yields than 1.5 g/L. However, no significant difference was seen between the use of POPOP and bis-MSB as a secondary wavelength shifter. Since Lumirror reflects well at the emission wavelength of PPO, we decided to use a PC+TMB+PPO (3 g/L) mixture for further studies.

\section{Preparation and Filling of the Bell Jar}
\label{belljarprep}
\subsection{Cleaning}
The Lumirror was cut into pieces that would line all surfaces of the bell jar. The pieces were cleaned with ethanol and placed in a vacuum to dry. The vacuum was later filled with nitrogen. The bell jar was scrubbed clean with a cleaning solution (1:50 ratio of Deteregent 8 to deionized water), and then rinsed with deionized water. Ethanol was repeatedly forced through a spray ball inserted into the top of the bell jar with high nitrogen pressure.  This process forcefully washed the surfaces of the bell jar and removed water. Nitrogen was left blowing through the bell jar for a couple days to further dry it out. We then filled the bell jar with an argon atmosphere while we opened it and installed the Lumirror. The bell jar was closed and dried with nitrogen for several days. The bell jar was then pumped down to a vacuum of 2.2$\times$10$^{-5}$ mbar and tested for leaks with a helium leak detector.

\subsection{Filling}
To fill the bell jar, we set up a distillation column 
for PC and TMB. The distillation column was connected to a large Erlenmeyer flask, which was in turn connected to a port in the bell jar. The whole system was airtight when the ports in the distillation column were closed, and it was held at a vacuum during the distillation. Before beginning the distillation, the amount of PPO needed to obtain a 3 g/L final concentration was added to the Erlenmeyer flask. PC and TMB were distilled into the flask until it was roughly halfway full. The flask was then shaken until the contents were well mixed, which were poured into the bell jar. We repeatedly distilled more PC and TMB this way until we had filled the bell jar with a total of 10 L of each.

\section{Filled Bell Jar Measurements}
\label{bjresultssect}
\subsection{Measurements}
\begin{figure}
 \centering
 \includegraphics[width=.9\linewidth]{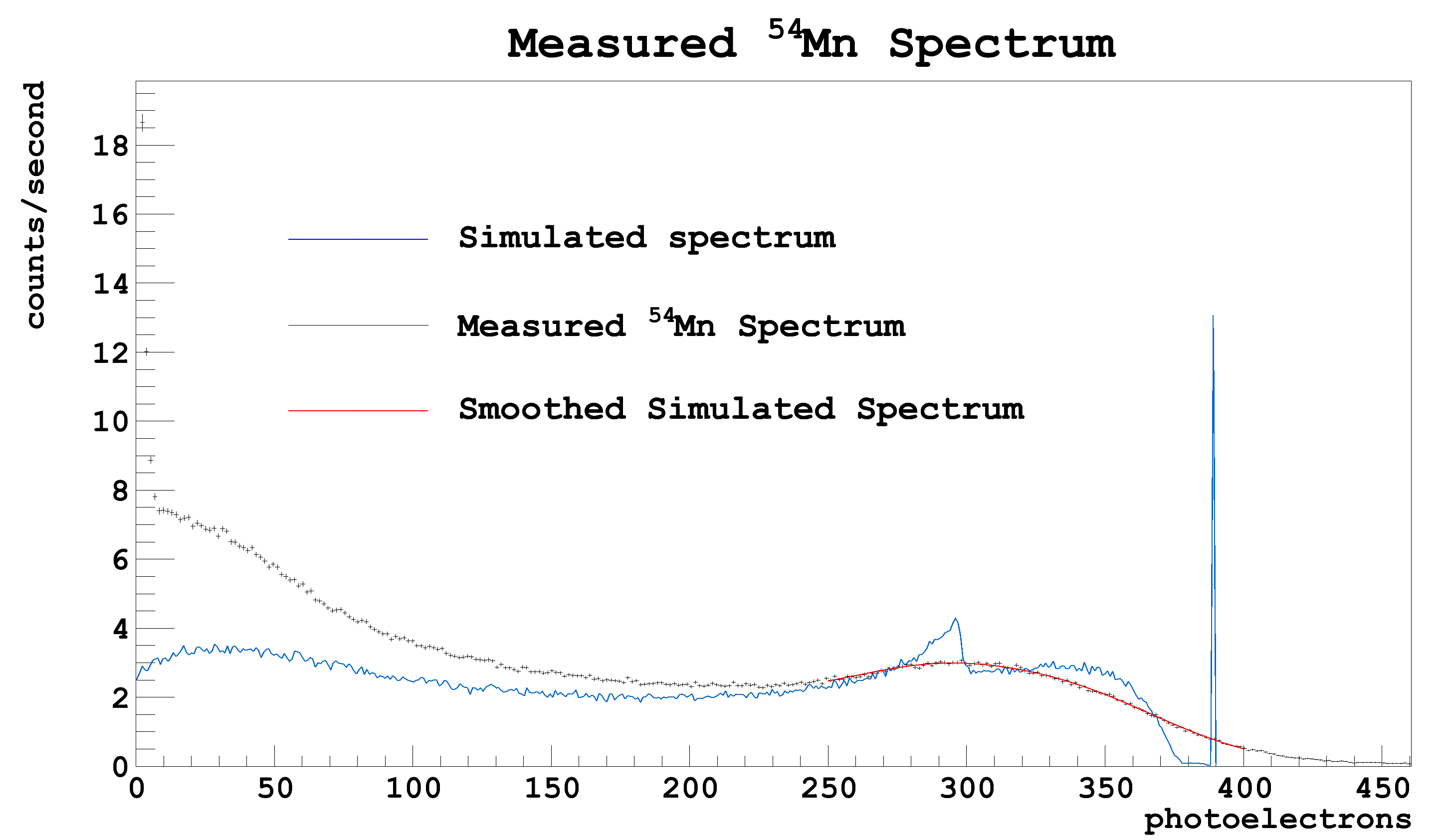}
 \caption{A sample $^{54}$Mn spectrum measured in the bell jar, background subtracted and scaled to the number of photoelectrons by the single photoelectron peak. A smoothed simulated spectrum has been fit to the data around the Compton edge (red); the blue shows the simulated spectrum before smoothing. The large narrow spike around 390 photoelectrons is the 835 keV full energy peak, while the broader spike just below 300 photoelectrons is the Compton edge. The part of the spectrum beyond the Compton edge is due to multiple scattering. The fit was performed specifically around the Compton edge in order to avoid effects from ionization quenching.}
 \label{comptonspecs}
\end{figure}

The light yield in the bell jar was measured over a span of 42 days.  The bell jar and source were kept inside a house of lead bricks to reduce background. Measurements were taken by by placing a 1 $\mu$Ci $^{54}$Mn $\gamma$ source (834.8 keV) on top of the bell jar near the center. We then removed the source and measured the background and the single photoelectron peak. The light yield was calculated by considering a GEANT4 simulation of the energy deposited in the bell jar from the same source. The background-subtracted $\gamma$ spectrum was then scaled by the single photolectron peak. The simulated $\gamma$ spectrum was convolved with a Gaussian response function which accounted for the resolution and light yield of the detector as fit parameters. This spectrum was then fit to the measured spectrum around the Compton edge (Fig.~\ref{comptonspecs}). 

\subsection{Results}
\begin{figure}
 \centering
 \includegraphics[width=.8\linewidth]{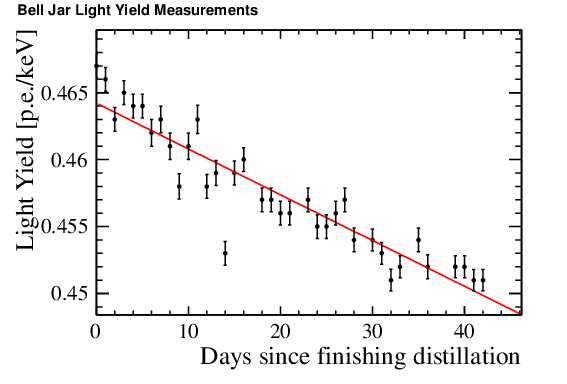}
 \caption{The light yield (in photoelectrons per keV) measured in the bell jar over 42 days, neglecting the exponential tail in the single photoelectron distribution, decreasing at a rate of 0.52$\pm 0.02$\% per week.}
 \label{bjlightyields}
\end{figure}
Immediately after filling the bell jar, we measured the light yield to be 0.466$\pm$0.001 p.e./keV.  However, the light yield was observed to steadily decrease over time, at an average rate of 0.52$\pm 0.02$\% per week (Fig.\ref{bjlightyields}). This may be due to the pseudocumene interacting with the metal oxides in the stainless steel walls of the bell jar. As can be seen in Fig.~\ref{pcinss}, prior observations have been made showing that pseudocumene's attenuation length decreases when it is exposed to stainless steel for a prolonged period of time~\cite{jayben}. It is difficult to do a direct comparison between the degradation rate observed in~\cite{jayben} and the rate observed here, since the studies in~\cite{jayben} used pure pseudocumene, and different stainless steel samples were used for both sets of measurements. Nevertheless, an order-of-magnitude estimate of the size of the effects shows a rough agreement between the two rates of degradation, lending some plausibility to this explanation.

\section{A Measurement of a Neutron Source with the Bell Jar}
\label{neutronsect}
The neutron spectrum was measured in the bell jar with an AmBe ($\alpha,n$) source.  The phototube was hooked up to a CAEN V1720 250 MHz digitizer.  We took three different runs: one background, one with a $^{54}$Mn source, and one with the AmBe source.   The $^{54}$Mn source was used to scale the horizontal axis to energy in keVee.   

\begin{figure}
\centering
\includegraphics[width=0.9\linewidth]{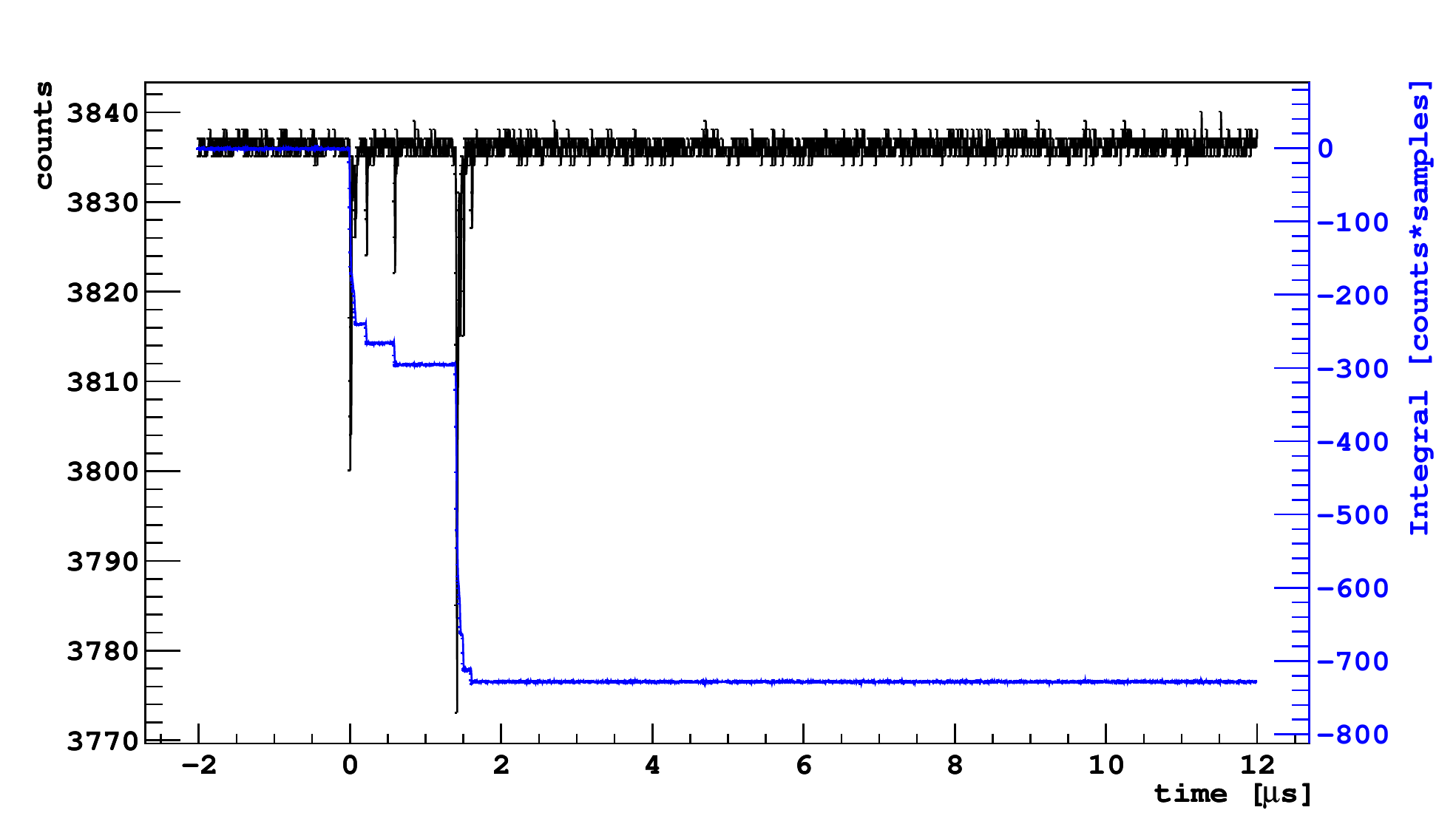}
\caption{An example ADC waveform of a neutron capturing on $^{10}$B in the bell jar. The thermalization signal can be seen around 0 $\mu$s, and the neutron capture signal is seen around 1.5 $\mu$s. The blue curve on the waveform shows the total integral of the curve from the start of the acquisition window to a given time.}
\label{ambencapturewf}
\end{figure}

For the AmBe run, we expect two detectable signals caused by neutrons.  A ``prompt" signal occurs when the neutrons are thermalized by elastically scattering off protons.  This is a fast process that occurs over tens of nanoseconds~\cite{ReinesNeutrons}.  We expect this prompt signal to be heavily quenched to $\sim$10\%~\cite{hong_scintillation_2002}.  We also expect a capture signal with a mean delay of approximately 3.3 $\mu$s after the thermalization signal~\cite{Wri2010}.  An example waveform of a neutron capturing on $^{10}$B can be seen in Fig.~\ref{ambencapturewf}. At 0 $\mu$s, one can see the thermalization signal, followed by the capture signal at $\sim1.5\mu$s.

This capture should produce a signal at or between 40 and 60 keVee from the lithium and the $\alpha$ in Equation \ref{reaction}, with a possible 478 keV $\gamma$.  We base our expectation for this first number on measurements from various commercial scintillators. The measurements taken with these commercial sources were performed by~\cite{ScintillatorS1,ScintillatorS2,ScintillatorT}, using either AmBe or PuBe neutron sources. However, these commercial scintillator cocktails were different from the liquid scintillator cocktail being considered here. ~\cite{ScintillatorS1} used Saint-Gobain BC523A commercial scintillator, ~\cite{ScintillatorS2} used Saint-Gobain BC523A2 and Eljen Technology EJ339A2 commercial scintillators, and~\cite{ScintillatorT} used a 50\% Toluene and 50\% TMB cocktail with 4 g/L PBD wavelength shifter and 20 mg/L POPOP secondary wavelength shifter. Since the optical properties, including the effects of ionization quenching, vary between scintillator cocktails, the strength of the $\alpha$ signal relative to the $\gamma$ signal can vary significantly between different studies. 

To capture both the prompt thermalization signal and the delayed capture signal in a single trigger ``event" we digitized 13 $\mu$s worth of data for each trigger.

In selecting signals for analysis we applied several cuts to our data.  First, we required that the phototube was not saturated during any pulse within the event.  Second, we required that at least two pulses were found within the event.  

\begin{figure}
\centering
\includegraphics[width=.8\textwidth]{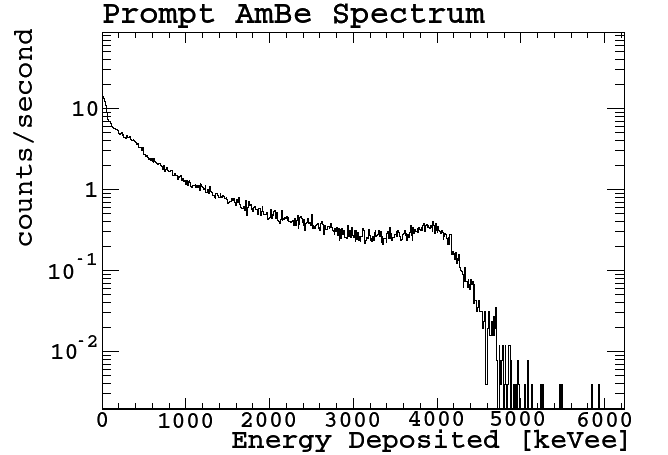}
\caption{The background-subtracted prompt energy spectrum from the AmBe source. The peak at high energy is the Compton edge of the 4.4 MeV $\gamma$ produced by the $^9$Be($\alpha$,n)$^{12}$C reaction with a branching ratio of ~64\%.}
\label{promptspec}
\end{figure}

\begin{figure}
\centering
\includegraphics[width=.6\textwidth]{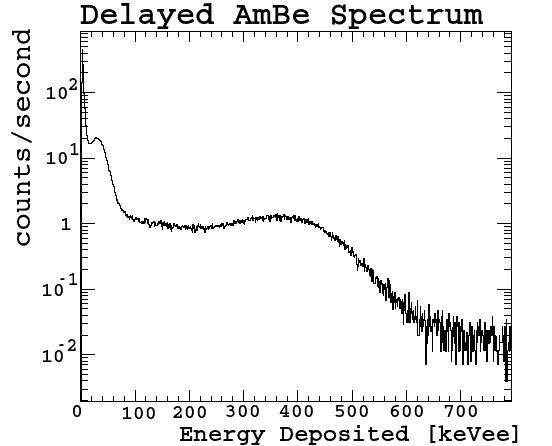}
\caption{The background-subtracted delayed energy spectrum from the AmBe source. The peak at ~30 keVee represents the energy deposit of the lithium and $\alpha$ alone, and the broad peak around 400 keVee represents the Compton spectrum of the $\gamma$ plus the lithium and the $\alpha$. }
\label{neutronsfinally}
\end{figure}

Once the events were selected, we made cuts on the individual pulses.   Individual pulses within an event were identified by looking for signals where the ADC went below a threshold of about a few photoelectrons in amplitude.  In this analysis, we required that peaks within 10 ns of each other be grouped under the same ``pulse" in order to group multiple scatters of the same neutron.  For the delayed spectrum, we set time cuts such that the pulses had to occur at least 0.65 $\mu$s after the trigger.  This time cut was done largely to avoid PMT after-pulsing, which peaks at 0.55 $\mu$s.  Because of the high amount of low-energy noise, we accepted only the largest energy pulse in this time window for our delayed spectrum.  After performing this analysis on both runs, we background-subtracted to produce Figs.~\ref{promptspec} and~\ref{neutronsfinally}, which show the prompt and delayed spectra, respectively. The prompt spectrum in Fig.~\ref{promptspec} is dominated by the Compton spectrum of the 4.4 MeV $\gamma$ produced by the $^9$Be($\alpha$,n)$^12$C reaction with a branching ratio of 64\%. The peak at high energies is the Compton edge, which is 4.2 MeV.

In Fig.~\ref{neutronsfinally} the broad peak at $\sim$400 keVee comes from the energy deposited by the $^7$Li, the $\alpha$, and Compton scatters from the $\gamma$ in Equation~\ref{reaction}.  The lower energy peak comes from the $\alpha$ and $^7$Li produced by this reaction, with contributions from the ground state reaction that do not produce the $\gamma$, as well as the excited state reaction when the $\gamma$ escapes the bell jar without leaving a signal. This peak occurs around 30 keVee, which was lower than was found by the aforementioned studies~\cite{ScintillatorS1,ScintillatorS2,ScintillatorT}.

This difference may be due to the different optical properties of the scintillators being compared. However, it is also possible that this difference is a result of the degradation of the scintillator due to it reacting with the stainless steel of the bell jar, causing nuclei to be more heavily quenched than electrons, since this measurement was taken several months after filling the bell jar. Another possibility is that the $\alpha+^7$Li peak is lower than expected may be due to the fact that the peak is dominated by the capture to the excited state, which has a branching ratio of $94\%$, where the $\gamma$ escapes, since the $\alpha$ and $^7$Li nucleus in this reaction are at a lower energy than in the case with no $\gamma$.

The effects of neutron and $\alpha$ quenching in the scintillator is the subject of a study that will be published in a separate paper.
  
\section{Optical Monte Carlo Simulation}
Two optical Monte Carlo simulations of the bell jar were performed, one using GEANT4-9.5.0, and the other using a custom-built simulation~\cite{emilythesis}. The custom-built simulation allows the user to generate packets of photons within a volume, representing scintillation events. It then tracks the photons as they reflect on different boundaries between materials. Attenuation of photons in the scintillator was modelled using an exponential probability distribution with the length of the photon's track between each step, assuming that the attenuation length was independent of the photons' wavelength. Before using this simulation to model the bell jar, it was first tested on the small cells discussed in Section~\ref{smallcellsect}, and the attenuation length was tuned based on these comparisons. It is worth noting that while the attenuation length was tuned, all other parameters in the custom-built simulation were separately measured as a function of wavelength, leaving the attenuation length as the only free parameter. These simulations agreed with the measurements given in Table~\ref{smallresults} to within 30\%.

The simulations include the stainless steel bell jar vessel, the Lumirror reflector with measured reflectance curve (modeled as a boundary between the bell jar and its contents), the PMT, and the quartz window around it. The quartz window is given an index of refraction of 1.45. The photocathode is given a quantum efficiency as reported by Hamamatsu~\cite{HamCat}. Additionally, the photocathode is given a diameter of 6.51 cm and the PMT is assumed to collect photoelectrons emitted from the cathode on the first dynode with an efficiency of 0.85. When a photon hits the photocathode, it is detected with a probability given by the quantum efficiency at its wavelength; if it was not detected, the photon is given a 20\% chance of reflecting off the photocathode, and is otherwise absorbed and not detected. The refractive index used for the scintillator is 1.43, the average of the refractive indices of PC and TMB.

Reflections off of the walls of the bell jar are pure Lambertian, since Lumirror is a primarily diffuse reflector. Reflections off the photocathode are treated as being entirely specular. The quartz is given a decadic extinction coefficient of 0.005 cm$^{-1}$, typical for quartz, to allow light to attenuate in the quartz window.

The scintillator's scintillation yield is assumed to be 12,670 photons/MeV. This number is based on the yield measured by Borexino in~\cite{Borex1997}, adjusted for the 50\% dilution by TMB and the increased yield from the extra PPO concentration, as measured in Section~\ref{smallcellsect}. Since energy deposited in the PC is quickly and efficiently transferred non-radiatively to the PPO, the scintillation spectrum used is the emission spectrum of PPO. Ionization quenching is calculated for the GEANT4 simulation using the default calculation handled by the scintillation module, with an assumed value of Birks's constant of 0.0115 cm/MeV~\cite{Borex1997}.

Light propagation is handled by using the measured values of PC and TMB  absorption spectra. At any given wavelength, the absoprtion length is calculated by assuming equal concentrations of PC and TMB and that the molar extinction coefficients of a mixture add linearly proportional to the components' concentration~\cite{Borex1997}. This leads to the equation
\begin{equation*}
 \frac{1}{\lambda_{\text{PC+TMB}}} = \frac{1}{2\lambda_{\text{PC}}} + \frac{1}{2\lambda_{\text{TMB}}}
\end{equation*}
Additional effects due to the absorption and re-emission by the PPO are accounted for by using the measured absorption length of PPO to determine if a photon gets absorbed by the PPO. In the case where PPO re-absorbs photons, the simulation produces another photon, drawn from the emission spectrum of PPO at a longer wavelength than the absorbed photon came in with, with an 82\% probability, corresponding to the quantum efficiency of PPO~\cite{borexjohnson}.

For these simulations, a $^{54}$Mn source is placed on the top of the bell jar and allowed to decay, producing a 835 keV $\gamma$ that is propagated. The total energy deposited into the scintillator is then calculated, as well as the total number of photoelectrons detected by the PMT. This ratio is used to determine the simulated light yield. 

Both simulations, the one performed in GEANT4 and the custom-made one, gave results consistent with each other. 

The primary uncertainty in this light yield measurement comes from the systematic uncertainty in the optical properties of the scintillator, especially the scintillation yield, for which the used value has $\sim$13\% uncertainty. This uncertainty comes from the uncertainty reported by Borexino in~\cite{Borex1997} compounded with the uncertainty in the the effects of the different PPO concentration, estimated in Section~\ref{smallcellsect}. Additional uncertainties from the other optical properties such as the attenuation length and the quartz reflectivity combine to give a total systematic uncertainty of $\sim$15\%.

The total light yield calculated from this simulation with the $^{54}$Mn source at the top of the bell jar is 0.46 p.e./keVee. This number is in very close agreement with the measured light yield of 0.466 p.e./keVee. 

Similar simulations were performed with the source at various positions with respect to the bell jar to test the position dependence of the scintillation. For these simulations, the maximum variation observed in the light yield is 8.6$\pm$3.9\%, compared to a maximum measured variation in the data of 3.3$\pm$0.4\%. 

Similar simulations have been performed for the DarkSide-50 neutron veto, a 4 meter diameter sphere with a roughly 1.2 m tall by 0.7 m diameter cryostat made of electropolished stainless steel and equipped with 110 Hamamatsu R5912 PMTs. These simulations use the same properties for the scintillator and reflector as used in the tests reported here. These simulations produce a 478 keV $\gamma$ (the energy of a $\gamma$ produced by a neutron capture on $^{10}$B) near the cryostat. The GEANT4 simulation predicts a light yield of 0.48$\pm0.09$ p.e./keV and the custom-made simulation predicts a light yield of 0.53$\pm$0.08 p.e./keV. These predictions agree very closely with the light yield of 0.54$\pm$0.04 p.e./keVee observed by the DarkSide-50 neutron veto with 50\% PC, 50\% TMB and $\sim$3 g/L PPO~\cite{ds50first}.

\section{Conclusions and Applications}
We have shown that through the creation of highly clean scintillator cells and a small optically-sealed bell jar setup, we could test the effectiveness of various scintillator cocktails and reflectors.  Through this work we discovered that Lumirror and a scintillator mix of equal parts PC and TMB with 3 g/L PPO would produce a good light collection for larger neutron detectors.  

We also created a small neutron veto prototype out of our bell jar, modifying it such that it could be filled with newly distilled scintillator.  We determined the light yield of this miniature detector to measure the effects of light attenuation in the scintillator.  Though the bell jar did show degradation in the light yield over time, our current understanding based on evidence in~\cite{jayben} is that this degradation is due to interactions between the pseudocumene and the stainless steel. In this case, we would expect the magnitude of this effect to be proportional to the vessel's surface-to-volume ratio. Since the bell jar had a surface-to-volume ratio of 21 m$^{-1}$ and the DarkSide-50 neutron veto has a ratio of 1.5 m$^{-1}$, we expect this effect to be much smaller in a full-sized neutron veto. Moreover, if substantial degradation is observed in the scintillator, it may be possible to recover the light yield by re-purifying the scintillator. We have since coated the inside of the bell jar with a protective ETFE coating to reduce this effect, although measurements have not yet been made.

We have also simulated the optics of the bell jar and found very good agreement between light yield predictions in the Monte Carlo and those observed in data, showing a strong ability to understand and predict the light collection efficiency of such a veto. GEANT4 simulations of the DarkSide-50 neutron veto have predicted a light yield of 0.48$\pm0.09$ p.e./keVee. 

A measurement of an AmBe source in the bell jar showed that this type of detector is capable of detecting neutrons. With a simple data acquisition system we were able to see peaks both from the $\gamma$ from the excited lithium decay, and also from the rarer lower-energy captures. 

This document has demonstrated that measurements of the light yield of the scintillator, the reflectance of the reflector, and the PMT quantum efficiencies are important parameters for determining the neutron rejection efficiency of a neutron veto for low background experiments. Furthermore, it can be seen from the measurements presented here that such a highly efficient veto is feasible with a design similar to the one described in this document. 

This design should be particularly applicable to dark matter experiments such as DarkSide-50~\cite{ds50first}, which has a light yield of $\sim$0.5 p.e./keV and predicts a neutron vetoing efficiency $\gtrsim 99.2\%$~\cite{ds50first,Agnes:2015_uar}, and others that are considering using such a veto. While the current SuperCDMS design does not call for a neutron veto, a similar design to the one presented here is being considered for future upgrades~\cite{calkins_prototyping_2015}. The LZ experiment is planning to use a neutron veto with a design that is more different from the one presented here~\cite{malling_after_2011,the_lz_collaboration_lux-zeplin_2015}. The design currently planned by the LZ collaboration uses linear-alkylbenzene, doped with gadolinium in the form of GdCl$_3$, and using PPO as a wavelength shifter. This veto design will consist of a set of acrylic vessels surrounding the liquid xenon TPC, which will contain the liquid scintillator and provide a 61 cm thick buffer, as well as a liquid xenon ``skin layer'' immediately outside the TPC, inside the cryostat. The LZ collaboration predicts that this design will have a light yield of 0.13 p.e./keV in the liquid scintillator, and that the combined skin layer and liquid scintillator will have a neutron vetoing efficiency of about 93--96\%~\cite{the_lz_collaboration_lux-zeplin_2015}.

\section{Acknowledgements}
We thank Aldo Ianni and Yury Suvorov for measuring the decrease in light yield observed when a pure PC scintillator is diluted 50\% with TMB. We also thank Matthias Laubenstein for measuring the radioactive contamination of samples of Lumirror foil, Allen Nelson for helping build the bell jar prototype, and Michael Souza blowing the glass for the liquid scintillator cells and helping seal them. We acknowledge support from the National Science Foundation (US Grant PHY-1103987, ``Particle Astrophysics at Princeton: Solar Neutrino and Dark Matter Studies with Borexino and DarkSide'').

\clearpage
\section{References}
\bibliographystyle{elsarticle-harv}
\bibliography{biblio}{}
\end{document}